\documentclass[english,reprint, aps, preprintnumbers,superscriptaddress]{revtex4-1}
\usepackage[T1]{fontenc}
\usepackage[latin9]{inputenc}
\usepackage[letterpaper]{geometry}
\usepackage{babel}
\usepackage{array}
\usepackage{units}
\usepackage{multirow}
\usepackage{amsmath}
\usepackage{amssymb}
\usepackage{graphicx}
\usepackage[unicode=true,pdfusetitle,
 bookmarks=true,bookmarksnumbered=false,bookmarksopen=false,
 breaklinks=false,pdfborder={0 0 1},backref=false,colorlinks=false]
 {hyperref}

\makeatletter

\providecommand{\tabularnewline}{\\}

\usepackage{scrextend}
\usepackage{hyperref}
\binoppenalty=\maxdimen
\relpenalty=\maxdimen

\makeatother

\begin{document}


\title{Benchmark neutrinoless double-beta decay matrix elements in a light
nucleus}

\affiliation{Department of Physics, Iowa State University, Ames, IA 50010, USA}
\affiliation{Department of Physics, University of North Carolina, Chapel Hill, NC 27514, USA}
\affiliation{FRIB/NSCL and Department of Physics and Astronomy, Michigan State University, East Lansing, MI 48824, USA}
\author{R.A.M.~Basili} \affiliation{Department of Physics, Iowa State University, Ames, IA 50010, USA}
\author{J.M.~Yao} \affiliation{Department of Physics, University of North Carolina, Chapel Hill, NC 27514, USA} \affiliation{FRIB/NSCL and Department of Physics and Astronomy, Michigan State University, East Lansing, MI 48824, USA}
\author{J.~Engel} \affiliation{Department of Physics, University of North Carolina, Chapel Hill, NC 27514, USA}
\author{H.~Hergert} \affiliation{FRIB/NSCL and Department of Physics and Astronomy, Michigan State University, East Lansing, MI 48824, USA}
\author{M.~Lockner} \affiliation{Department of Physics, Iowa State University, Ames, IA 50010, USA}
\author{P.~Maris} \affiliation{Department of Physics, Iowa State University, Ames, IA 50010, USA}
\author{J.P.~Vary} \affiliation{Department of Physics, Iowa State University, Ames, IA 50010, USA}
\vskip 0.25cm

\date{\today}
\begin{abstract}
We compute nuclear matrix elements of neutrinoless double-beta decay mediated by light Majorana-neutrino exchange in the $A=6$ system. The goal is to benchmark two
many-body approaches, the No-Core Shell Model and the Multi-Reference In-Medium
Similarity Renormalization Group. We use the SRG-evolved chiral N3LO-EM500 potential for the nuclear interaction, and make the approximation that isospin is conserved. We compare the results of the two approaches as a function of the cutoff on the many-body basis space. Although differences are seen in the predicted nuclear radii, the ground-state energies and neutrinoless double-beta decay matrix elements produced by the two approaches show significant agreement. We discuss the implications for calculations in heavier nuclei. 
\end{abstract}


\maketitle
\section{introduction}
\label{sec:level1}
Since the discovery of the lepton flavor violation in neutrino oscillations
\citep{Ahmad:2001an,Eguchi:2002dm,Fukuda:1998mi}, identifying whether
the neutrino is a Majorana fermion (i.e.~its own antiparticle) has
become a priority in nuclear and particle physics. However, because
neutrinos are charge neutral and nearly massless, they are notoriously
difficult to detect, and their properties remain only partly understood.
Major theoretical and experimental collaborative efforts are already
underway to study neutrino properties \citep{Martin-Albo:2015rhw,Albert:2014awa,Gilliss:2018lke,KamLAND-Zen:2016pfg,Alfonso:2015wka,Agostini:2017iyd,Aalseth:2017btx,Andringa:2015tza,Iwata:2016cxn,Cirigliano:2017djv,Contessi:2017rww,Coraggio:2017bqn,Jiao:2017opc,Tiburzi:2017iux,Shanahan:2017bgi,Horoi:2017gmj}.
Determining whether neutrinos are indeed Majorana particles would not
only shed light on the mechanism behind neutrino mass generation,
but would also provide insight on leptogenesis and the universe's
apparent matter-antimatter asymmetry. 

Neutrinoless double-beta ($0\nu\beta\beta$) decay is a hypothetical
lepton-number-violating (LNV) nuclear transition where two neutrons
decay to two protons and two electrons but no anti-neutrinos (or the
reverse with leptons exchanged with their antiparticles). Observing
$0\nu\beta\beta$ decay would confirm the existence of a LNV process,
and is commonly viewed as the best means of learning whether neutrinos
are Majorana particles. Experiments designed to detect $0\nu\beta\beta$ decay
in ton-scale amounts of $^{76}\text{Ge}$, $^{136}\text{Xe}$, and
other materials have already put impressive limits on the $0\nu\beta\beta$-decay
half-life \citep{Gilliss:2018lke,KamLAND-Zen:2016pfg,Agostini:2017iyd},
and these limits will only become more accurate as additional data
is collected. For a more complete description of current and past
efforts as well as some of the underlying theory, see Refs.$\ $\citealp{Cremonesi:2013vla,DellOro:2016tmg,Engel:2016xgb,GomezCadenas:2011it,Henning:2016fad}
and references therein.

While of enormous significance in itself, the experimental detection
or non-detection of $0\nu\beta\beta$ decay will be insufficient to
pin down or put limits on extra-Standard-Model parameters such as
the average neutrino mass. Because the decay rate depends on the $0\nu\beta\beta$-decay
nuclear matrix elements (NMEs), interpreting the experimental results
requires the accurate calculation of those NMEs. However, at present
the calculated NMEs in the heavy nuclei of interest differ by a factor
of two to three \citep{Engel:2016xgb}. In addition, calculated NMEs
for $\beta$ decay are usually smaller than experimental values, and
the reasons for these differences are only now being understood in
a quantitative way \citep{Gysbers:2019uyb}. To shed light on these
differences, it is helpful to examine weak processes in light nuclei,
where calculations are better controlled than in the heavy nuclei
we must eventually grapple with. Thus, while not viable
for $0\nu\beta\beta$-decay experiments, light nuclei are a practical option for
benchmarking.

The purpose of this study is to calculate the NMEs (for $0\nu\beta\beta$ decay
mediated by light Majorana-neutrino exchange) in the $A=6$ system.
Benchmarking different many-body methods and identifying important
features that affect the NMEs in these light nuclei will both test
the approaches that we will apply in heavy nuclei and help us anticipate
issues that may arise there. Assessing the convergence behavior of
the decay NMEs with increasing model-space size is of particular importance,
as it will help quantify uncertainties in heavier nuclei where more
severe basis truncation is computationally required. Thus, we consider
the ground-state-to-ground-state $0\nu\beta\beta$ decay of $^{6}\text{He}\rightarrow{}^{6}\text{Be}$,
which, while kinematically disallowed, involves the same decay operator
that determines the allowed decay rates in heavy nuclei. 

We employ two \textit{ab initio} many-body approaches: the No-Core
Shell Model (NCSM) and the Multi-Reference In-Medium Similarity Renormalization
Group (MR-IMSRG). The NCSM is a large-scale diagonalization method
that yields exact results in the limit of an infinitely large configuration
space. On the other hand, the MR-IMSRG (a variation of the IMSRG in
which the method's reference state contains explicitly built-in correlations)
yields approximate solutions to the many-body Schr{\"o}dinger equation
within a systematically improvable truncation scheme. That is, where
the NCSM includes all many-body correlations up to the given basis
cutoff by construction, the MR-IMSRG only includes many-body correlations
up to a cutoff in the many-body expansion. In exchange, the computational
effort of the MR-IMSRG scales much more favorably with particle number
and configuration space size, which makes it capable of modeling both
light and heavy nuclei. While both methods treat all nucleons as active,
they can also be used to generate effective interactions and operators
for traditional Shell-model calculations in heavier nuclei \citep{Bogner:2014baa,Jansen:2014qxa,Jansen:2015ngw,Stroberg:2015ymf,Stroberg:2016ung,Dikmen:2015tla,Barrett:2017ovf}.

For both the MR-IMSRG and NCSM calculations performed in this work,
we assume good isospin symmetry to facilitate the comparison of their
results, though it should be noted that we could drop this assumption
at the cost of introducing more complex methods \citep{Song:2017ktj,Yao:2018qjv}.
For both approaches we adopt the next-to-next-to-next-to-leading chiral
order (N3LO) Entem-Machleidt two-body potential with regulator cutoff
$\Lambda=500$ MeV (referred to as 'N3LO-EM500') \citep{Entem:2003ft,Machleidt:2011zz},
to model the nucleon-nucleon (NN) interaction. The potential is expressed
in the harmonic oscillator (HO) basis with energy scale $\hbar\Omega=20$
MeV, and softened by SRG evolution to the scale of $\lambda=2.0$
$\text{fm}^{-1}$ (with the relative kinetic energy, $T_{rel}$, as
the generator \citep{Bogner:2009bt}) prior to many-body calculations.

Our examination of the $A=6$ system with the NCSM is similar to the
studies in Refs.$\ $\citealp{Cockrell:2012vd,Shin:2016poa}, but
differs from both in: the NN-interaction used, the extrapolations
employed, our focus on $0\nu\beta\beta$ decay, and our comparison
with the MR-IMSRG approach. Our study also offers a point of comparison
to the computation of $0\nu\beta\beta$-decay NMEs arising from an
array of LNV mechanisms in light nuclei by using \textit{ab initio}
Variational Monte-Carlo (VMC) techniques \citep{Pastore:2017ofx},
though our study is distinguished by our use of a different NN-interaction
and our focus solely on $0\nu\beta\beta$ decay mediated by light
Majorana-neutrino exchange.

The rest of this paper is structured as follows: Section$\ $\ref{sec:level2}
briefly outlines the derivation of the $0\nu\beta\beta$-decay operator
as defined in Refs.$\ $\citealp{Engel:2016xgb,Avignone:2007fu,Simkovic:1999re}.
We provide a brief review of the NCSM in \ref{sec:level3-1}, and
the MR-IMSRG in \ref{sec:level3-2}. Section$\ $\ref{sec:level4}
compares the ground-state energy and square radius (in \ref{sec:level4-1}),
and analyzes the contributions to the total $0\nu\beta\beta$-decay
NME (in \ref{sec:level4-2}). Finally, Section$\ $\ref{sec:level5}
reviews our findings and concludes the discussion. Additional details
regarding our extrapolation methods and tables of calculated values
are provided in 
the \hyperref[sec:appendix]{appendix}.

\section{\label{sec:level2}$0\nu\beta\beta$ Decay with Light Majorana-Neutrinos}

\begin{figure}
\mbox{
\includegraphics[width=8.6cm]{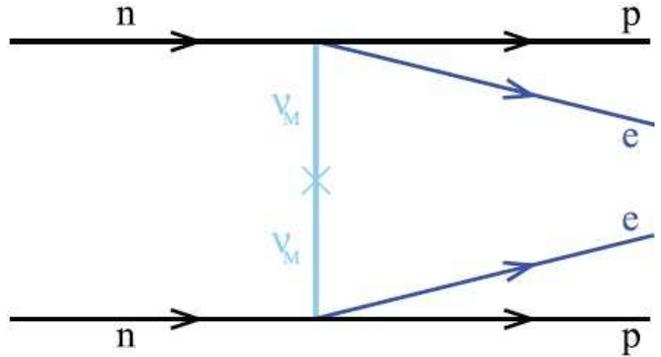}
}\caption{\label{fig:II.1} Feynman diagram (modified from Ref.$\ $\citealp{Engel:2016xgb})
for $0\nu\beta\beta$ decay mediated by light-neutrino exchange. Two
neutrons (n) decay into two protons (p), emitting two electrons ($\text{e}^{-}$).
No neutrinos are emitted, implying that they are Majorana particles
($\nu_{\text{M}}$).}
\end{figure}
We consider $0\nu\beta\beta$ decay caused by the exchange of the
three light Majorana neutrinos and the Standard-Model weak interaction
as depicted in Fig.$\ $(\ref{fig:II.1}); all contributions from
other LNV processes are neglected. 

Drawing on Refs.$\ $\citealp{Engel:2016xgb,Avignone:2007fu} and
the approximations employed there, we write the $0\nu\beta\beta$-decay
rate as 
\begin{equation}
\left[T_{1/2}^{0\nu}\right]^{-1}=G_{0\nu}\left(Q,Z\right)\left|M_{0\nu}\right|^{2}\left|\underset{k}{\sum}m_{k}U_{ek}^{2}\right|^{2}\,,\label{eq:II.2}
\end{equation}
where $Q$ is the difference between initial ($i$) and final ($f$)
state energies, (i.e. $Q\equiv E_{i}-E_{f}$), $Z$ is the proton
number of the final nucleus, $m_{k}$ is the Majorana mass eigenvalue,
and $U_{ek}$ is the element of the neutrino mixing matrix that connects the electron
neutrino with mass eigenstate $k$. $G_{0\nu}\left(Q,Z\right)$
comes from the phase-space integral, which has been evaluated with
improved precision in Refs.$\ $\citealp{Kotila:2012zza,Stoica:2013lka}. 

In this study, we focus on the $^{6}\text{He}\rightarrow{}^{6}\text{Be}$
ground-state-to-ground-state NME, $M_{0\nu}$ \citep{Simkovic:1999re,Horoi:2009gz,Rodin:2006yk},
obtained from the $0\nu\beta\beta$-decay many-body operator, $O_{0\nu}$,
as
\begin{equation}
M_{0\nu}=\langle^{6}\text{Be}|O_{0\nu}|^{6}\text{He}\rangle.\label{eq:II.3 benchmark ME}
\end{equation}
Our notation follows that of Ref. \citealp{Suhonen:2007zza} unless
specified otherwise.

\subsection{\label{sec:level3-2-1}The $0\nu\beta\beta$-decay Matrix Elements}

The many-body operator $O_{0\nu}$ is conventionally divided into
three contributions, labeled Fermi, Gamow-Teller (GT), and tensor.
We use the symbol $O$ to generically denote any one of these contributions'
corresponding two-body operator, which may always be written in second-quantized
form as

\begin{flalign}
O & =\frac{1}{4}\underset{\alpha\beta\gamma\delta}{\sum}\langle\alpha\beta|O|\gamma\delta\rangle a_{\alpha}^{\dagger}a_{\beta}^{\dagger}a_{\delta}a_{\gamma}\label{eq: II.4 general op}
\end{flalign}
where $a^{\dagger}$ and $a$ create and annihilate nucleons, respectively,
in single-particle states. A given single-particle state $\alpha$
is defined by the quantum numbers $n_{\alpha}$, $l_{\alpha}$, $s_{\alpha}$,
$j_{\alpha}$, $t_{\alpha}$, $m_{j\alpha}$, and $m_{t\alpha}$,
which correspond to the radial, angular momentum, spin, total angular
momentum, isospin, angular momentum projection, and isospin projection,
respectively. Greek indices $\alpha,\beta,\gamma,\delta$ are used
to denote single-particle states, while the corresponding Roman indices
$a,b,c,d$ refer to the reduced set of quantum numbers, such that
$a_{\alpha}^{\dagger}\equiv a_{a,m_{j\alpha},m_{t\alpha}}^{\dagger}$.
We define spherical tensor/isotensor versions of the annihilation
operators as 
\begin{flalign}
\hat{a}_{\delta} & \equiv\left(-1\right)^{j_{\delta}+m_{j\delta}+\frac{1}{2}+m_{t\delta}}a_{d,-m_{j\delta},-m_{t\delta}}\,,
\end{flalign}
such that{\small{}
\begin{equation}
a_{\delta}a_{\gamma}=\left(-1\right)^{j_{\gamma}+j_{\delta}+m_{j\gamma}+m_{j\delta}+1}\hat{a}_{c,-m_{j\gamma},-m_{t\gamma}}\hat{a}_{d,-m_{j\delta},-m_{t\delta}}\,.
\end{equation}
}{\small \par}

For the ground-state-to-ground-state portion of the $^{6}\text{He}\rightarrow^{6}\text{Be}$ transition,
we may narrow our scope to components of the two-body operators that
contribute to $0^{+}\rightarrow0^{+}$ NMEs. Expanding Eq.$\ $(\ref{eq: II.4 general op})
into doubly-reduced tensorial components in the $JT$-coupled two-body
isospin representation yields for this transition 
\begin{flalign}
O_{0,-2}^{0,2} & =-\frac{1}{4\sqrt{3}}\underset{abcd}{\sum}\underset{J}{\sum}\left(\mathcal{N}_{ab}\left(J,1\right)\mathcal{N}_{cd}\left(J,1\right)\right)^{-1}\nonumber \\
 & \cdot(ab;J\,1|||O^{0,2}|||cd;J\,1)\nonumber \\
 & \cdot\left[\left[a_{a}^{\dagger}a_{b}^{\dagger}\right]^{J,1}\left[\hat{a}_{c}\hat{a}_{d}\right]^{J,1}\right]_{0,-2}^{0,2}\,,\label{eq:II.8 0nbb op}
\end{flalign}
where brackets denote tensor products with tensor, isotensor couplings
in superscripts and their corresponding projections in subscripts,
$\mathcal{N}_{ij}(J,T)\equiv\sqrt{1-\delta_{ij}(-1)^{J+T}}/(1+\delta_{ij})$
is an antisymmetrization factor, and the triple lines '|||' denote
doubly-reduced two-body matrix elements (TBMEs). In Eq.$\ $(\ref{eq:II.8 0nbb op})
we implicitly include only two-body states that satisfy the Pauli
exclusion principle in the sum over nucleon states (or, effectively,
we only consider values of $i$, $j$, $J$, and $T$ such that $\mathcal{N}_{ij}(J,T)\neq0$).

We express the total NME ($M_{0\nu}$) as the sum of the Fermi ($M_{0\nu}^{F}$),
GT ($M_{0\nu}^{GT}$), and tensor ($M_{0\nu}^{T}$) contributions
\begin{equation}
M_{0\nu}=M_{0\nu}^{F}+M_{0\nu}^{GT}+M_{0\nu}^{T}\,.\label{eq:II.10 comps}
\end{equation}
These three NME contributions are developed for the many-body initial
and final nuclear states from the doubly-reduced TBMEs of the three
corresponding two-body operators. We evaluate the NMEs by summing the two-body contribution from each unique pair of the system's
nucleons. We calculate the TBMEs with the two-body operators{\small{}
\begin{align}
O_{0\nu}^{F}\left(r\right) & =\frac{4R}{\pi g_{A}^{2}}\int_{0}^{\infty}|\mathbf{q}|d|\mathbf{q}|\frac{j_{0}(|\mathbf{q}|r)h_{F}(|\mathbf{q}|)}{|\mathbf{q}|+\bar{E}-(E_{i}+E_{f})/2}\tau_{1}^{+}\tau_{2}^{+}\,,\nonumber \\
O_{0\nu}^{GT}\left(r\right) & =\frac{4R}{\pi g_{A}^{2}}\int_{0}^{\infty}|\mathbf{q}|d|\mathbf{q}|\frac{j_{0}(|\mathbf{q}|r)h_{GT}(|\mathbf{q}|)\boldsymbol{\sigma}_{1}\cdot\boldsymbol{\sigma}_{2}}{|\mathbf{q}|+\bar{E}-(E_{i}+E_{f})/2}\tau_{1}^{+}\tau_{2}^{+}\,,\nonumber \\
O_{0\nu}^{T}\left(r\right) & =\frac{4R}{\pi g_{A}^{2}}\int_{0}^{\infty}|\mathbf{q}|d|\mathbf{q}|\frac{j_{2}(|\mathbf{q}|r)h_{T}(|\mathbf{q}|)\boldsymbol{S}_{12}}{|\mathbf{q}|+\bar{E}-(E_{i}+E_{f})/2}\tau_{1}^{+}\tau_{2}^{+}\,,\label{eq:II.11}
\end{align}
}where $\boldsymbol{q}$ is the momentum transfer, $r=|\boldsymbol{r}_{1}-\boldsymbol{r}_{2}|$
is the magnitude of the inter-nucleon position vector, and $\hat{\boldsymbol{r}}$
is the corresponding unit vector. Additionally, $\boldsymbol{r}_{1\backslash2}$,
$\boldsymbol{\sigma}_{1\backslash2}$, and $\tau_{1\backslash2}^{+}$
respectively denote the labeled nucleon's position operator, spin
operator, and isospin raising operator (transforming neutrons to protons),
while $\boldsymbol{S}_{12}=3\boldsymbol{\sigma}_{1}\cdot\hat{\boldsymbol{r}}\boldsymbol{\sigma}_{2}\cdot\hat{\boldsymbol{r}}-\boldsymbol{\sigma}_{1}\cdot\boldsymbol{\sigma}_{2}$
is the tensor operator. The NMEs contain $r$-dependence through the
spherical Bessel functions $j_{0}$ and $j_{2}$ in Eq.$\ $(\ref{eq:II.11}),
and, for several heavy parent nuclei, have been shown to vanish at
small distances $r$, fall off like $\nicefrac{1}{r}$ at large distance,
and have a typical range of a few femtometers (fm) \citep{Simkovic:2007vu}.
Hence, we expect good convergence with the basis space for these operators
in our calculations.

The neutrino potentials, $h$, are defined in momentum space as {\small{}
\begin{align}
h_{F}\left(|\boldsymbol{q}|\right) & \equiv-g_{V}^{2}(\boldsymbol{q}^{2})\,,\nonumber \\
h_{GT}\left(|\boldsymbol{q}|\right) & \equiv g_{A}^{2}(\boldsymbol{q}^{2})-\frac{g_{A}(\boldsymbol{q}^{2})g_{P}(\boldsymbol{q}^{2})\boldsymbol{q}^{2}}{3m_{N}}+\frac{g_{P}^{2}(\boldsymbol{q}^{2})\boldsymbol{q}^{4}}{12m_{N}^{2}}+\frac{g_{M}^{2}(\boldsymbol{q}^{2})\boldsymbol{q}^{2}}{6m_{N}^{2}}\,,\nonumber \\
h_{T}\left(|\boldsymbol{q}|\right) & \equiv\frac{g_{A}(\boldsymbol{q}^{2})g_{P}(\boldsymbol{q}^{2})\boldsymbol{q}^{2}}{3m_{N}}-\frac{g_{P}^{2}(\boldsymbol{q}^{2})\boldsymbol{q}^{4}}{12m_{N}^{2}}+\frac{g_{M}^{2}(\boldsymbol{q}^{2})\boldsymbol{q}^{2}}{12m_{N}^{2}}\,,
\end{align}
}where $g_{M}\left(q^{2}\right)=\left(1+\kappa_{1}\right)g_{V}\left(q^{2}\right)\simeq4.706g_{V}\left(q^{2}\right)$
(with the anomalous nucleon isovector magnetic moment $\kappa_{1}=3.706$),
and the Goldberger\textendash Treiman relation $g_{P}\left(q^{2}\right)=2m_{N}g_{A}\left(q^{2}\right)/\left(\boldsymbol{q}^{2}+m_{\pi}^{2}\right)$
(with nucleon mass $m_{N}$ and pion mass $m_{\pi}$) connects the
pseudoscalar and axial terms \citep{Simkovic:2009pp,Engel:2016xgb}.
The conservation of the vector current implies that $g_{V}\equiv\left.g_{V}\left(q^{2}\right)\right|_{0}=1$,
while the value $g_{A}\equiv\left.g_{A}\left(q^{2}\right)\right|_{0}\simeq1.27$
may be extracted from neutron $\beta$-decay measurements. Their momentum
transfer dependence is $g_{V}\left(q^{2}\right)=g_{V}\left(1+q^{2}/\Lambda_{V}^{2}\right)^{-2}$
and $g_{A}\left(q^{2}\right)=g_{A}\left(1+q^{2}/\Lambda_{A}^{2}\right)^{-2}$
where $\Lambda_{V}=850$ MeV and $\Lambda_{A}=1040$ MeV are the vector
and axial masses, respectively. The nuclear radius $R=1.2A^{\frac{1}{3}}\approx2.2$
fm is inserted by convention to make the matrix elements dimensionless,
with a compensating factor absorbed into $G_{0\nu}$ in Eq.$\ $(\ref{eq:II.2}).
Finally, $\bar{E}$ is an estimate of the average intermediate-state
energy, the choice of which has been shown to have only a mild influence
on the decay amplitude \citep{Avignone:2007fu}. We employ the value
$\bar{E}-(E_{i}+E_{f})/2\equiv5\,\text{MeV}$ throughout this work.

In other prescriptions, the operators defined by Eq.$\ $(\ref{eq:II.11})
are sometimes multiplied by an additional radial function, $f(r)$,
designed to take into account short-range correlations that are omitted
by Hilbert-space truncations performed in the many-body calculations
\citep{Miller:1975hu,Roth:2005pd,Brueckner:1955zza,Benhar:2014cka,Muther:2000qx}.
In this work, we assume all relevant nucleon-nucleon correlations
are embedded in the many-body wavefunctions generated in our NCSM
and MR-IMSRG model spaces and employ no additional radial function.
We numerically integrate the inner products of the operators in Eq.$\ $(\ref{eq:II.11})
using relative HO states to obtain reduced matrix elements in the
relative basis. These elements are then converted to M-scheme TBMEs
via a Moshinsky transformation \citep{Tobocman:1981yao,Barrett:2013nh}
before being employed in many-body calculations.

\subsection{\label{sec:level3-2-2}$0\nu\beta\beta$ decay in $^{6}\text{\textbf{He}}$
with Isospin Symmetry}

When considering isovector operators, a common challenge shared by
many $ab\,initio$ nuclear approaches (particularly those relying
on finite matrix methods) arises when the initial and final nuclei
are not the same, as the many-body spaces for the two will generally
differ. In NCSM calculations, this problem usually requires the many-body
eigenstate wavefunctions of the two systems to be calculated independently.
In the MR-IMSRG, two different unitary transformation operators must
be constructed, one for the initial nucleus and one for the final
nucleus. 

While solutions to this problem have been developed
for the NCSM and have been implemented for the MR-IMSRG \citep{Yao:2018qjv},
a careful choice of transition can circumvent the issue when isospin
conservation is a good approximation. Thus, we assume that isospin
symmetry is obeyed in the mirror nuclei $^{6}$He and $^{6}$Be.

The ground states of $^{6}{\rm Be}$ and $^{6}{\rm He}$ are characterized
by total angular momentum $\mathcal{J}=0$ and isospin $\mathcal{T}=1$,
with projections $\mathcal{T}_{z}=-1,+1$, respectively. If isospin
symmetry is obeyed, the two-body density of Eq.$\ $(\ref{eq:II.8 0nbb op})
may be rewritten in terms of the $^{6}\text{He}$ two-body density
alone as
\begin{flalign}
 & \langle^{6}\text{Be}|\left[\left[a_{a}^{\dagger}a_{b}^{\dagger}\right]^{J,1}\left[\hat{a}_{c}\hat{a}_{d}\right]^{J,1}\right]_{0,-2}^{0,2}|^{6}\text{He}\rangle\nonumber \\
 & =\sqrt{6}\langle^{6}\text{He}|\left[\left[a_{a}^{\dagger}a_{b}^{\dagger}\right]^{J,1}\left[\hat{a}_{c}\hat{a}_{d}\right]^{J,1}\right]_{0,0}^{0,2}|^{6}\text{He}\rangle.\label{eq:II.14 iso sym}
\end{flalign}

\section{\label{sec:level3}benchmarked methods}

Both the NCSM and MR-IMSRG can provide accurate results when applied
in light nuclei. The MR-IMSRG has the advantage that, with suitable
approximations, it can be applied in heavier systems \citep{Bogner:2014baa,Stroberg:2015ymf,Stroberg:2016ung}.
For the NCSM one may envision applications in heavier systems by merging
it with renormalization approaches or by introducing an inert core
and deriving effective interactions for valence-space Shell model
calculations (see e.g., Refs. \citealp{Jansen:2014qxa,Dikmen:2015tla,Jansen:2015ngw,Barrett:2017ovf}).
The approximations involved in these envisioned approaches to heavier
nuclei will also require benchmarking.

Both methods consider the $A$-body nuclear Hamiltonian, $H$, consisting
of a relative kinetic-energy term and interaction terms, i.e.
\begin{equation}
H=\frac{1}{2Am_{N}}\underset{i<j}{\overset{A}{\sum}}\left(p_{i}-p_{j}\right)^{2}+V_{NN}+V_{NNN}+...
\end{equation}
where $m_{N}$ is the average nucleon mass, $V_{NN}$ is the NN-interaction,
and $p_{i}$ denotes the momentum of nucleon $i$. We follow the convention
for two-body operators where summations over nucleon pairs are performed
under the ordering given by $i<j$ to avoid counting the same pair
twice. The term $V_{NNN}$ denotes three-body interactions, also called
3-nucleon forces (3NFs), which may be supplemented by higher-body interactions.
Although studies have demonstrated that 3NFs can have a significant
impact on calculated nuclear observables \citep{Barrett:2013nh},
their inclusion would greatly increase computational cost and is thus
deferred to future efforts. We therefore consider here only the NN-interactions
from N3LO-EM500 \citep{Entem:2003ft,Machleidt:2011zz}, which is charge-dependent.

\subsection{\label{sec:level3-1}No-Core Shell Model}

The NCSM \citep{Barrett:2013nh} is a configuration-interaction (CI)
approach in which the many-body basis states, $|\Phi\rangle$, are
expressed as Slater determinants of single-particle states occupied
by the system's nucleons, or 
\begin{equation}
|\Phi\rangle=\mathcal{\mathcal{A}}\left[\underset{i}{\prod}|\phi_{\alpha_{i}}\rangle\right]\,,
\end{equation}
where $|\phi_{\alpha_{i}}\rangle$ denotes a single-particle state
with quantum numbers $\alpha_{i}$ occupied by nucleon $i$, and $\mathcal{A}$
is an antisymmetrization operator that carries both the sign permutations
of the determinant and an overall normalization factor. Our
NCSM approach features separate Slater determinants for the neutrons
and protons, and the resulting many-body basis is specific to the
nucleus under consideration. For a given application, we form total
Slater determinants of fixed parity and fixed total angular momentum
projection $M_{J}$.

The infinite HO basis (with energy scale fixed by the usual parameter
$\hbar\Omega$) is the conventional choice of single-particle basis
and is used in this work. Additional details on the HO basis functions
may be found in Ref.$\ $\citealp{Barrett:2013nh}. 

The nuclear many-body wavefunctions, $\Psi\left(r_{1},...,r_{A}\right)$,
satisfy the A-body Schr{\"o}dinger equation and are obtained by solving
the Hamiltonian matrix eigenvalue problem 
\begin{equation}
H|\Psi\rangle=E|\Psi\rangle\label{eq: III.A.4}
\end{equation}
where $E$ is the eigenenergy of nuclear state $|\Psi\rangle$. Beginning
with the kinetic-energy and interaction TBMEs in the HO basis, one
constructs the $A$-body Hamiltonian matrix elements in the many-body
basis as $\langle\Phi_{\mu}|H_{A}|\Phi_{\nu}\rangle$, where the indices
$\mu$ and $\nu$ label the many-body basis states. The many-body
eigenstates are then linear combinations of many-body basis states:
\begin{equation}
|\Psi\rangle=\underset{\mu}{\overset{\infty}{\sum}}c_{\mu}|\Phi_{\mu}\rangle
\end{equation}
where $c_{\mu}$ are the normalized coefficients of the many-body
basis states $|\Phi_{\mu}\rangle$. For practical calculations, the
infinite many-body basis requires truncation, which one controls by
using a basis cutoff parameter. For NCSM calculations performed in
this study, we employ the cutoff parameter $N_{\text{max}}$, which
denotes the maximum number of HO excitation quanta allowed in the
many-body basis above the minimum number required by the Pauli principle
\citep{Barrett:2013nh}. 

Solving Eq.$\ $(\ref{eq: III.A.4}) with the resulting finite many-body
Hamiltonian then becomes a large (but generally sparse) matrix eigenvalue
problem. We obtain the solution with the hybrid OpenMP/MPI CI code
Many Fermion Dynamics for nucleons (MFDn). The code is optimized for
solving the large sparse matrix eigenvalue problem by using a Lanczos-like
algorithm to determine the desired lowest-lying energy eigenvalues
and corresponding eigenvectors. The eigenvectors are then used with
other operator matrix elements to calculate that operator's expectation
values during post-processing. For more details on MFDn, see Refs.$\ $\citealp{MARIS201097,Aktulga:2014mfd,DBLP:journals/corr/ShaoAYNMV16}. 

By solving the system in a sequence of increasingly large bases, one
can extrapolate to the result when using the complete basis (i.e.
when the matrix dimension of $H$ goes to infinity and the calculation
becomes exact). Any other observable can also, in principle, be extrapolated
to this limit, and such extrapolations are a distinguishing feature
of No-Core Full-Configuration (NCFC) studies \citep{Maris:2008ax}. 

\subsection{\label{sec:level3-2}Multi-Reference In-Medium Similarity Renormalization
Group}

Here we provide a brief overview of the MR-IMSRG; a more complete
description may be found in Refs.$\ $\citealp{Hergert:2015awm,Hergert:2016etg,Hergert:2018wmx}.
For an initial Hamiltonian $H$, the flow equation 
\begin{equation}
\dfrac{dH(s)}{ds}=[\eta(s),H(s)]\,,\label{eq:III.B.1 flow}
\end{equation}
determines a unitary transformation of the Hamiltonian. Here $\eta$
is called the generator of scale transformations and $s$ is the flow
parameter, defined such that $H(s)\mid_{s=0}$ is just $H$. The ground-state
energy is simply given by the expectation value of the evolved Hamiltonian
$H(s)$ in the reference state. Instead of solving the set of differential
equations for $H(s)$ in Eq.$\ $(\ref{eq:II.10 comps}), one can
solve a similar flow equation for the unitary transformation operator
$U(s)$, 
\[
\frac{dU\left(s\right)}{ds}=\eta\left(s\right)U\left(s\right)\,,
\]
whose solution can formally be written in terms of the ${\cal S}$-ordered
exponential 
\begin{equation}
U\left(s\right)=\mathcal{S}\,\text{exp}\int_{0}^{s}ds'\eta\left(s'\right)\,,
\end{equation}
which is short-hand for the Dyson series expansion of $U(s)$. As
shown first by Magnus, it is possible to rewrite the unitary transformation
operator as $U(s)\equiv e^{\Omega(s)}$, a step that transforms the
equation for $U(s)$ into one for $\Omega$ \citep{Morris:2015yna}:
\begin{equation}
\frac{d\Omega\left(s\right)}{ds}=\underset{n=0}{\overset{\infty}{\sum}}\frac{B_{n}}{n!}\left[\Omega\left(s\right),\eta\left(s\right)\right]^{\left(n\right)}\,.
\end{equation}
The nested commutators in this equation are given by\begin{subequations}
\begin{align}
\left[\Omega\left(s\right),\eta\left(s\right)\right]^{\left(0\right)} & =\eta\left(s\right)\,,\\
\left[\Omega\left(s\right),\eta\left(s\right)\right]^{\left(n\right)} & =\left[\Omega\left(s\right),\left[\Omega\left(s\right),\eta\left(s\right)\right]^{\left(n-1\right)}\right]\,,
\end{align}
\end{subequations} and $B_{n=0,1,2,\cdots}$ are the Bernoulli numbers
$\{1,-1/2,1/6,\cdots\}$.

The expectation value of any operator $O$ is then given by $\langle\Phi|O\left(s\right)|\Phi\rangle=\langle\Phi|e^{\Omega(s)}O{}^{-\Omega(s)}|\Phi\rangle$,
and can be evaluated with the Baker-Campbell-Hausdorff formula: 
\begin{equation}
e^{\Omega(s)}Oe^{-\Omega(s)}=\underset{n=0}{\overset{\infty}{\sum}}\frac{1}{n!}\left[\Omega\left(s\right),O\right]^{\left(n\right)}.
\end{equation}

In the MR-IMSRG calculations performed here, we express all operators
in normal-ordered form with respect to a reference state $|\Phi\rangle$
in order to control the proliferation of induced terms. We keep up
to normal-ordered two-body operators throughout the calculation, in
accordance with the MR-IMSRG(2) truncation described in Ref.$\ $\citealp{Hergert:2018wmx}.
We use particle-number-projected HFB quasiparticle vacua as reference
states, and adopt the Brillouin generator \citep{Hergert:2016etg}.
We numerically solve the flow equation for values of $s$ large enough
so that the solutions are very close to their asymptotic limits. The
underlying Hamiltonian that defines both the projected HFB reference
state and the starting point for the flow equation is determined by
using the same TBMEs in the single-particle HO basis that are used
in our NCSM calculations. However, unlike the NCSM, the MR-IMSRG is
formulated in the natural orbital basis of the reference state. Since
the reference state results from a projected HFB calculation in a
HO basis, the MR-IMSRG effectively explores a configuration space
controlled by the cutoff parameter $e_{\text{max}}$, which denotes
the maximum number of energy quanta that the HO components of any
natural orbital can have. In effect, for a given cutoff $e_{\text{max}}$,
the MR-IMSRG many-body basis will include single-particle excitations
up to $e_{\text{max}}$ (i.e. one-particle-one-hole, or 1p1h), two-particle
excitations (i.e. 1p1h+1p1h or 2p2h) up to $2e_{\text{max}}$, uncorrelated
three-body excitations (i.e. 1p1h + 1p1h + 1p1h or 1p1h + 2p2h) up
to $3e_{\text{max}}$, and so on. 

\section{\label{sec:level4}results and discussion}

Here we discuss the results of the NCSM and MR-IMSRG calculations.
We provide graphical representations of the results, as functions
of the basis cutoff parameters, to analyze the convergence of the
operators at the chosen basis scale of $\hbar\Omega=20$ MeV. Throughout,
we use solid dots to represent NCSM results and open boxes to represent
MR-IMSRG results. Similarly, we use solid lines to denote extrapolations
of the NCSM results and dashed lines to denote extrapolations of the
MR-IMSRG results. 

In order to compare the convergence behavior of results from the NCSM
and MR-IMSRG, we must consider the differences in their truncation schemes. We recall that the NCSM's cutoff parameter $N_{\text{max}}$
denotes the total number of allowed $excitation$ quanta in the system,
and $e_{\text{max}}$ denotes the maximum number of allowed energy
quanta possessed by any single nucleon. Since in $^{6}\text{He }$
at a given $N_{\text{max}}$ the highest number of quanta possessed
by any single-particle state will be $N_{\text{max}}+1$, we equate
the two cutoffs with the assignment $e_{\text{max}}\equiv N_{\text{max}}+1$
for our comparison. While this assignment is not exact, it ensures
that for a given pair of matched cutoffs, we use identical single-particle
bases under both truncation schemes.
Moreover, our use of this assignment to compare the results does not
preclude their examination from other perspectives. Instead, we merely
offer this assignment as a reasonable vehicle to present our results graphically.

We extrapolate our results to obtain predictions of observables at
the continuum limit and to better examine their convergence behavior;
the functional forms and other details of these extrapolations
are provided in
the \hyperref[sec:appendix]{appendix}.
We extrapolate our
NCSM and MR-IMSRG results for energy and square radii with formulae (Eq.$\ $(\ref{eq:0nbb ext.})
and Eq.$\ $(\ref{eq:rms R ext.}), respectively) inspired by those
provided in Refs.$\ $\citealp{Furnstahl:2012qg,Maris:2008ax}. Although these extrapolations were originally designed with the $N_{\text{max}}$ truncation scheme in mind, there is good reason from a theoretical perspective
to expect that the same extrapolation forms effective for the NCSM will
be effective for the results of IMSRG calculations \citep{Hergert:2015awm}.
Meanwhile,
as is the case for many nonscalar operator observables (with the exception of those
in significant investigations on extrapolating E2 observables \citep{Odell:2015xlw}),
precision extrapolation approaches for $0\nu\beta\beta$-decay observables
remain largely unexplored. Guided by the similarities of the observable's
$r$-dependence seen in Ref. \citep{Simkovic:2007vu} to that of nuclear
interactions, we employ the same simple exponential form applied for
the energy to extrapolate the $0\nu\beta\beta$-decay contributions. 
While we acknowledge that a thorough investigation of extrapolating $0\nu\beta\beta$-decay
NMEs is warranted for refined predictions and accurate uncertainty
estimates, we find that this form provides an adequate fit and proves sufficient
for this comparative study.

To facilitate our discussion of convergence, we refer to the speed
(with respect to the cutoff parameter) at which an eigenvalue result
approaches its asymptotic value as the result's ``convergence rate''. 
We gauge the convergence rate with the value of $N_{\text{max}}$
($e_{\text{max}}$) at which the extrapolation is within $5\%$ of
its value at the continuum limit, and denote this generally non-integer
value $\widetilde{N}_{5\%}$ ($\widetilde{e}_{5\%}$). 
%
%
%
Though this
metric relies heavily on the validity of the extrapolation, it provides
a functional estimate for both the relative convergence speeds between
results and the cutoffs required for reaching well-converged
values.

Finally, in the interest of understanding what the differences between
the extrapolated results of the two $ab\,initio$ calculations signify,
we briefly consider the general $A$-body system. For such a system,
the untruncated MR-IMSRG calculation would include all many-body correlations,
and would therefore provide identical results (within numerical noise)
as the NCSM at the continuum limit. By performing only the MR-IMSRG(2)
calculation, we expect the two approaches' results to converge to
different values that depend on how significant the neglected three-body
(up to $A$-body) correlations are to the observable in question.
Thus, beyond the mild uncertainty introduced by the extrapolation,
differences between the extrapolated results estimate the significance
of many-body correlations neglected by the MR-IMSRG(2) calculation. 

\subsection{\label{sec:level4-1}Ground-State Energy and Nuclear Square Radius}

\begin{figure*}[t]
\mbox{%
\includegraphics[width=17.2cm]{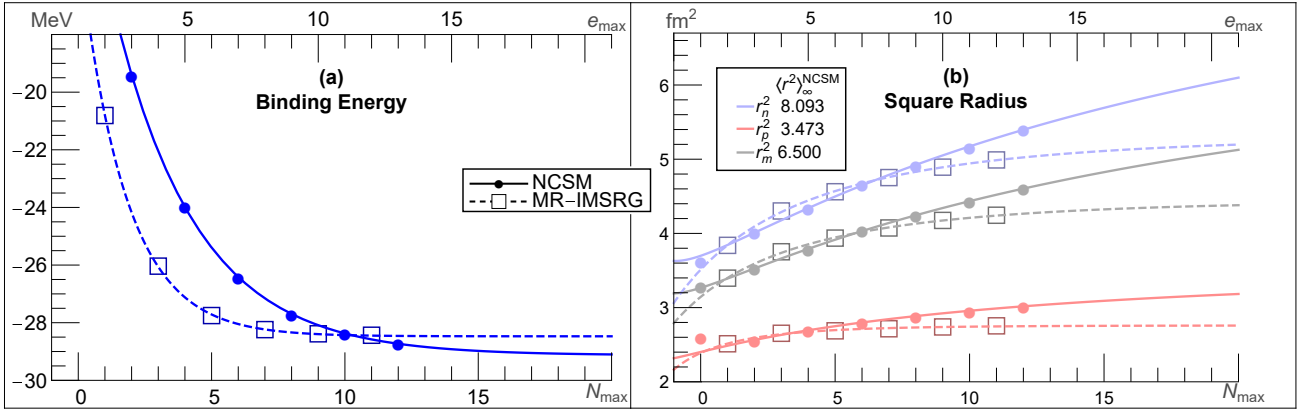}%
}\caption{\label{fig:1} Ground-State energy (a) and neutron (light blue), proton
(light red), and matter (gray) square radii ($r_{n}^{2}$, $r_{p}^{2}$,
and $r_{m}^{2}$ respectively) (b) of $^{6}\text{He}$ with varying
basis cutoff parameter from NCSM (solid circle) and MR-IMSRG(2) (open
square) \textit{ab initio} calculations. Solid and dashed lines denote
the NCSM and MR-IMSRG extrapolations, respectively. The realistic
N3LO-EM500 potential with energy scale $\hbar\Omega=20$ MeV and SRG
evolution scale $\lambda=2.0$ $\text{fm}^{-1}$ is used in all cases.
The asymptotic fit parameter, $\langle r^{2}\rangle_{\infty}$, of
the NCSM square radius extrapolations are listed in the legend of
(b) (see Eq.$\ $(\ref{eq:rms R ext.}) for extrapolation definition).
Fit parameters and plotted values are listed in Table$\ $(\ref{table:1})
for energy, and in Table$\ $(\ref{table:2}) for the square radii.}
\end{figure*}

The initial system is the $^{6}\text{He}$ nucleus in its ground state.
The calculated ground-state energy and neutron, proton, and matter
square radii ($r_{n}^{2}$, $r_{p}^{2}$, and $r_{m}^{2}$ respectively)
varying with basis truncation are shown in panels (a) and (b) of Fig.$\ $(\ref{fig:1})
respectively.

The NCSM ground-state energy extrapolation has converged to within
$5\%$ of its asymptotic value of -29.132 MeV by $\widetilde{N}_{5\%}\sim7.9$.
The MR-IMSRG(2) extrapolated energy converges somewhat faster by comparison,
with $\widetilde{e}_{5\%}\sim4.9$ and the asymptotic value of -28.472
MeV (around 2.3\% higher than the NCSM result).
%
%
Both extrapolated ground-state energies are underbound compared to
the experimental result of -29.272 MeV \citep{Riisager:1989hv}, as well as the extrapolated results of
-30.0(1) MeV and -29.87 MeV from two similar (but independent) NCFC
calculations of the $^{6}\text{He}$ ground-state \citep{Bogner:2007rx,Furnstahl:2012qg}
that used only the charge-independent parts of our strong-interaction
Hamiltonian.

Performing the same calculations for $^4\text{He}$ yields extrapolated binding energies of 28.305 and 28.316 MeV for the NCSM and MR-IMSRG(2), respectively. Subtracted from our $^6\text{He}$ binding energy results, this corresponds to 2$n$-separation energies of 0.827 MeV and 0.156 MeV. The experimental 2n-separation energy of $^6\text{He}$, by comparison, is approximately 0.975 MeV \citep{Wang:2017}.
%
The difference between these results is most likely a consequence of the truncations inherent to the MR-IMSRG(2). While the method probes a larger space of 2p2h excitations than the NCSM for a given pair of matched $N_{\text{max}}$ and $e_{\text{max}}$ single-particle bases, the MR-IMSRG(2) misses correlation energy from 3p3h and higher excitations that are included in the NCSM. We expect such correlations to play a more important role in a nucleus with a complex structure, like $^6\text{He}$, than in a compact nucleus like $^4\text{He}$.
We will analyze this issue in more detail using improved MR-IMSRG truncations \citep{Hergert:2018wmx} in the future.

Comparing the convergence rates of the square radii results between approaches, we see the MR-IMSRG(2) (NCSM) square radii consistently
converge faster (slower), with $\widetilde{e}_{5\%}\sim14,\,3.5,\,12$
($\widetilde{N}_{5\%}\sim72,\,30,\,63$) for neutron, proton, and
matter square radii, respectively. This large difference in convergence
speed primarily results from the use of natural orbitals in the
MR-IMSRG. However, we observe the MR-IMSRG(2) results converge
to a roughly $34\%,$ $21\%,$ and $31\%$ smaller value than the
NCSM results for the corresponding square radii.
All radii share the
slower convergence rate relative to the ground-state energy that is
commonly associated with the $r^{2}$ operator; a consequence of coming
from an effective operator with sensitivity to correlations outside the
characteristic length scale of the chosen HO basis \citep{Barrett:2013nh,Cockrell:2012vd,Furnstahl:2013vda,Furnstahl:2014hca,Coon:2012ab}.
In the NCSM (and to a lesser degree MR-IMSRG), the slow convergence
reflects the basis regularization of the infrared (IR) momentum region from
the HO basis truncation. In effect, because the basis's length scale
is chosen to favor convergence in energy, the basis requires higher
cutoffs to fully capture the longer-range correlations of the $r^{2}$
operator. The significantly faster convergence speed of the proton
square radius (compared to those of the neutron and matter square
radii) is a consequence of this effect, as the protons predominantly
remain in the core of the $^{6}$He ground-state halo structure \citep{Riisager:2012it}.
That is, since the protons are only found in the four-nucleon core,
the proton square radius operator correlations primarily exist
at the shorter distances pertinent to the core, and are thus better
encompassed by the scales of the chosen basis.

The benefit of the MR-IMSRG's renormalization can be seen in the improved
convergence observed in its results. In essence, the renormalization
decouples the NN-correlations existing outside the scales encompassed
by the basis, and distributes those correlations inside those scales.
The drawback is that some induced many-body forces must be neglected
in the process, an approximation that would explain the notable differences
seen in the extrapolated square radii. 
In the case of a light nucleus such as $^6\text{He}$ where a convergence trend can be established, we expect the NCSM extrapolated radii to be more accurate than those of the MR-IMSRG(2) for the given potential. 
Consequently, we conjecture that the smaller
MR-IMSRG(2) square radii reflect meaningful induced many-body correlations
that are being lost through the MR-IMSRG(2) many-body truncation.
Specifically, the fast convergence of the MR-IMSRG(2) results suggests
that the 1p1h and 2p2h correlations relevant to $r^{2}$ are well-accounted
for by $e_{\text{max}}=12$ and 24, respectively, and that the remaining
differences with the NCSM results are from higher many-body correlations
omitted by the MR-IMSRG(2) approach. 

\subsection{\label{sec:level4-2}$0\nu\beta\beta$-decay Matrix Element}

\begin{figure*}[t]
\mbox{
\includegraphics[width=17.2cm]{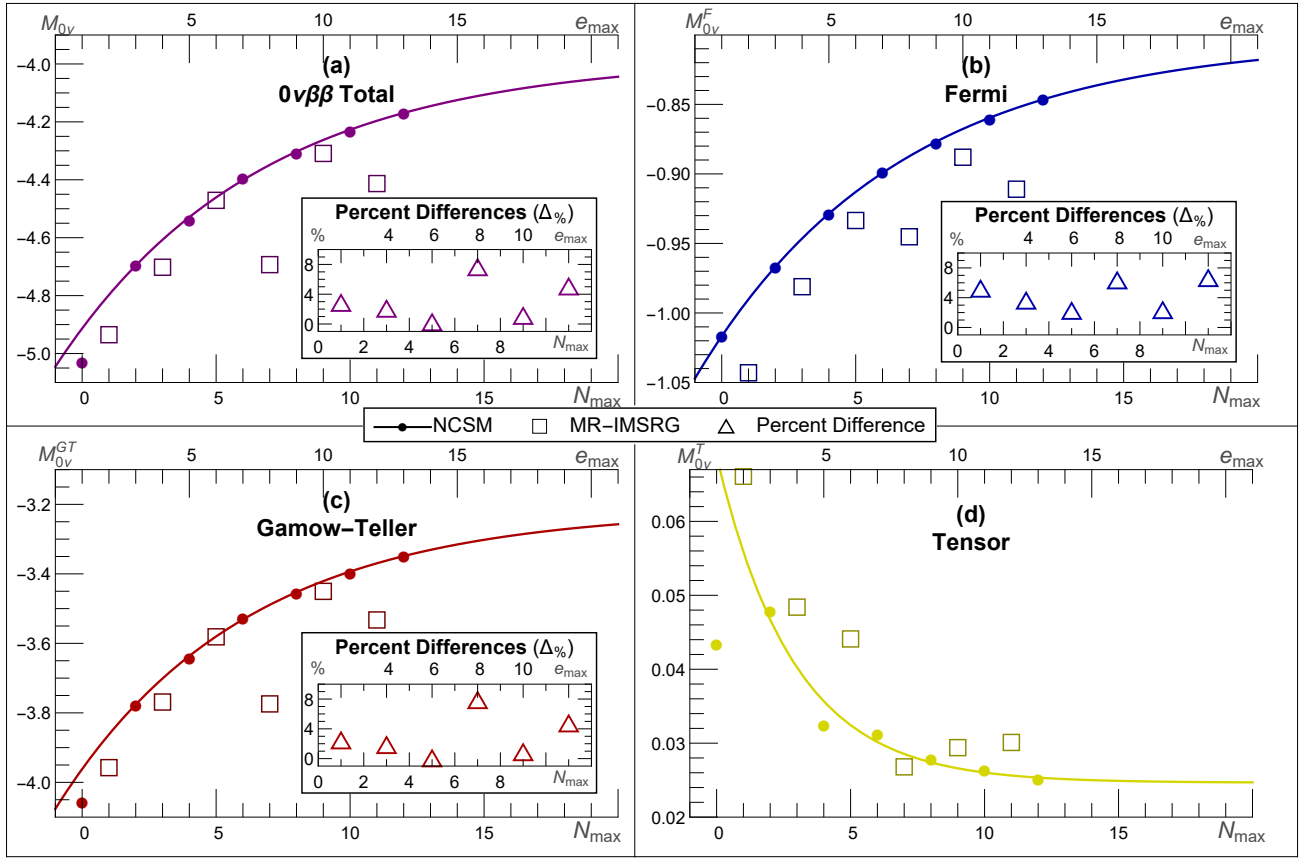}
}\caption{\label{fig:2} Ground-state-to-ground-state $0\nu\beta\beta$-decay
NME (a) for $^{6}\text{He}\rightarrow{}^{6}\text{Be}$ , decomposed
into its Fermi (b), GT (c), and tensor (d) contributions, as a function
of the basis cutoff for NCSM (solid circle) and MR-IMSRG(2) (open
square) \textit{ab initio} calculations. Solid lines denote NCSM extrapolations.
Each vertical axis is expanded for visibility. The three contributions
add to the total as specified by Eq.$\ $(\ref{eq:II.10 comps})).
Insets provide the percent difference in magnitude (open triangles)
between the MR-IMSRG results and the NCSM extrapolation as described
in the text. Plotted values and fit parameters are listed in Table$\ $(\ref{table:1}).}
\end{figure*}

We turn finally to the ground-state-to-ground-state $^{6}\text{He}\rightarrow{}^{6}\text{Be}$
$0\nu\beta\beta$-decay NME. As already mentioned, we assume isospin
symmetry so that the initial and final state are described by the
same wavefunction (except for an interchange of protons and neutrons).
We present our results in Fig.$\ $(\ref{fig:2}), where we recall
that discrete points represent results of many-body calculations while
lines represent fits specified by Eq.$\ $(\ref{eq:0nbb ext.}). We
decompose the total NME in (a) into its Fermi, GT, and tensor
contributions from Eq.$\ $(\ref{eq:II.10 comps}) in panels (b),
(c) and (d), respectively. Insets provide estimates for the percent
difference, $\Delta_{\%}$, between results of the two methods within
our mapping of their basis truncation schemes (we omit such estimates
for the numerically less significant tensor contribution). For a given
$e_{\text{max}}$, we calculate these values as 
\begin{equation}
\left(\Delta_{\%}\right)_{e_{\text{max}}}=200\left|\frac{\left(M_{0\nu}^{\text{IMSRG}}\right)_{e_{\text{max}}}-f\left(\left[N_{\text{max}}\right]_{e_{\text{max}}}\right)}{\left(M_{0\nu}^{\text{IMSRG}}\right)_{e_{\text{max}}}+f\left(\left[N_{\text{max}}\right]_{e_{\text{max}}}\right)}\right|\,,
\end{equation}
where $\left(M_{0\nu}^{\text{IMSRG}}\right)_{e_{\text{max}}}$ is
the NME result of the MR-IMSRG(2) calculation with cutoff $e_{\text{max}}$,
and $f\left(N_{\text{max}}\right)$ is the NCSM fit described by Eq.$\ $(\ref{eq:0nbb ext.})
evaluated at the mapped cutoff value $\left[N_{\text{max}}\right]_{e_{\text{max}}}=e_{\text{max}}-1$
(visualized in the figure by the intersection of a vertical line between
each open square and the NCSM extrapolation). While, much like the
mapping between cutoffs, these estimates require some level of arbitration,
we nevertheless find them a reasonable and useful tool for gauging
the differences between methods.

The $0\nu\beta\beta$-decay NMEs from the NCSM and MR-IMSRG approaches
agree remarkably well. Although the results of the MR-IMSRG(2) calculations
reflect significantly larger fluctuations with each step in the basis
cutoff, those fluctuations consistently remain less than a few percent
of the converged value, and the overall trends remain quite similar
to those of the NCSM. More importantly, the contributions, especially
the larger Fermi and GT contributions, show excellent agreement between
approaches.

The relative magnitudes of the contributions agree between approaches.
The GT contribution is around four times greater than the Fermi contribution,
while the tensor contribution is roughly two orders of magnitude smaller
and of opposite sign. The Fermi, GT, tensor, and total NME results
have $\widetilde{N}_{5\%}\sim12.6,\,10.8,\,10.3,\,11.7$, respectively,
which suggests only slightly slower convergence than that of the energy
but still significantly faster convergence than that of the NCSM square radii.

The MR-IMSRG $0\nu\beta\beta$-decay results resemble a saw-tooth
pattern for results beyond $e_{\text{max}}=4$ that gradually decreases
in magnitude as $e_{\text{max}}$ increases. The maximum deviation
of this pattern occurs in the GT contribution and reaches the order
of a few percent. The deviations of the tensor contribution appear
less systematic, though this may be a consequence of the contribution's
relatively small magnitude. The deviations in the Fermi and GT results
of the MR-IMSRG share a sign and are most visible at $e_{\text{max}}=8$,
where they consistently deviate in the negative direction. 

A mildly similar (though less pronounced) saw-tooth pattern is observed
in the tensor contribution of the NCSM results at the lowest $N_{\text{max}}$
cutoffs. Within NCSM calculations, such patterns (sometimes called
``odd-even effects'') are generally the consequence of alternating
signs in the asymptotic tails of the HO basis wavefunctions that are
introduced with each increment in $N_{\text{max}}$ \citep{Vary:2018jxg}.
In such cases, as the tail region of the calculated wavefunction shifts
with each increment, the tail begins to overlap a region of phase space in which
the effective operator is particularly active (i.e., has dominant
correlations). If the span of that active region is long enough to
require multiple steps in $N_{\text{max}}$ for the tail to pass through,
the result is a visible contribution to the observable that alternates
in sign. Naturally, the pattern disappears as $N_{\text{max}}$ increases
enough so that the effective operator's range is completely encompassed
by that of the basis.

One might wonder if the pattern observed in the MR-IMSRG
$0\nu\beta\beta$-decay results reflect a similar effect. However,
considering our MR-IMSRG(2) calculations employ natural orbitals and
not HO wavefunctions, the pattern's similarity may be entirely circumstantial.
Determining the origin of these deviations in the MR-IMSRG results
will require further study.

Despite these fluctuations making it somewhat challenging to make more
than qualitative observations, the trends of the NCSM and MR-IMSRG
results are remarkably similar. Indeed, the differences in the asymptotic
limits of the square radii in Fig.$\ $(\ref{fig:1}) do not appear
indicative of similar differences in the $0\nu\beta\beta$-decay NME
results. Similarly however, the more rapid convergence observed in
the MR-IMSRG(2) ground-state energy and square radii compared to that
of the NCSM does not appear to translate into a more rapid convergence
of the $0\nu\beta\beta$-decay NMEs in Fig.$\ $(\ref{fig:2}). The
differences between the two approaches' $0\nu\beta\beta$-decay results
appear to be of similar magnitudes as the saw-tooth deviations present
in the MR-IMSRG(2) results, and remain less than $5\%$ of the total
$0\nu\beta\beta$-decay NME at the maximum basis cutoff employed for
each method. 

Comparing our extrapolated $0\nu\beta\beta$-decay NMEs to those
calculated in the VMC approach with 3N correlations included \citep{Pastore:2017ofx},
we see that the magnitudes of both the GT and Fermi contributions
agree to within about $15\%$, while those of the tensor contribution
agree to within about $10\%$. For all three contributions, the VMC
results are larger. These differences may suggest a modest correction
from 3N correlations, though other differences between our study and
that of Ref.$\ $\citealp{Pastore:2017ofx} may play a significant role as well.

\section{\label{sec:level5}conclusion}

We find significant agreement between the NCSM and MR-IMSRG results
in our investigation of $0\nu\beta\beta$ decay in the $A=6$ system.
The difference in the calculated ground-state energy is only about
$\sim2.3\%$. We see measurable differences in the square-radius results
that offer an estimate for the effects of correlations that are omitted
by the MR-IMSRG(2) truncation at the normal-ordered two-body level.
It is interesting that these differences do not extend to the $0\nu\beta\beta$-decay
NMEs, which are remarkably similar in the two approaches, differing
by only $\sim4.3\%$ in the total NME at the largest basis cutoffs
considered. The convergence rate of the $0\nu\beta\beta$-decay NMEs
appears to be comparable to that of the energies.

The GT contribution dominates the $0\nu\beta\beta$-decay NME, comprising
$\sim80\%$ of its total. The Fermi contribution makes up most of
the remainder, and the tensor contribution is roughly two orders of
magnitude smaller and of opposite sign. 

Our estimates of the differences in the total $0\nu\beta\beta$-decay
NME between the two approaches do not exceed $9\%$ for any of the
basis cutoffs considered. Fluctuations in the MR-IMSRG results could
pose a minor obstacle for extrapolation, though their consistent saw-tooth
appearance may suggest these fluctuations are systematically correctable.
Beyond these fluctuations, the two approaches result in qualitatively
similar convergence for the $0\nu\beta\beta$-decay NME. 

The agreement between the two approaches for $0\nu\beta\beta$-decay
NMEs is encouraging, and warrants additional benchmarking. Unlike the transition studied in this work, the physically realistic 
$0\nu\beta\beta$-decay transition contributions do not possess a uniform sign as
a function of the pair separation \citep{WANG:2019}. That is, the 
$(0+,T)\rightarrow (0+,T-2)$ transitions of experimental interest \citep{Gilliss:2018lke,KamLAND-Zen:2016pfg,Agostini:2017iyd} have a node in the transition density,
making them much more sensitive to both short- and long-range correlations in the wavefunctions.
This sensitivity has been recently explored in another benchmark study comparing VMC and shell-model calculations of $0\nu\beta\beta$ decay in $A=10$
and $A=12$ systems, and has been seen to generate differences ranging anywhere from $30\%$ to $400\%$ between approaches \citep{WANG:2019}.  A similar comparison between the MR-IMSRG and NCSM approaches, including full isospin dependence, would provide a more stringent test of the many-body methods, and greater insight into problems that may appear when modeling the decay in heavier nuclei. 
The results observed here warrant such an investigation, and lend support to the application of MR-IMSRG to $0\nu\beta\beta$ decay in heavier nuclei \citep{Yao:2019rck}, 
where it is computationally more feasible than the NCSM.  
The good agreement between the two approaches for $0\nu\beta\beta$-decay NMEs is a promising development.

\begin{acknowledgments}
We offer special thanks to Roland Wirth for his help in validating the results of this work.
We acknowledge fruitful discussions with 
Sofia Quaglioni, Peter Gysbers,
Soham Pal, Shiplu Sarker, and Weijie Du. This work was supported in
part by the US Department of Energy (DOE), Office of Science, under
Grant Nos. DE-FG02-87ER40371, DE-SC0018223 (SciDAC-4/NUCLEI), DE-SC0015376
(DOE Topical Collaboration in Nuclear Theory for Double-Beta Decay
and Fundamental Symmetries), DE-SC0017887, and DE-FG02-97ER41019.
Computational resources were provided by the National Energy Research
Scientific Computing Center (NERSC), which is supported by the US
DOE Office of Science under Contract No. DE-AC02-05CH11231.
\end{acknowledgments}

\appendix
\section*{\label{sec:appendix}appendix: extrapolation methods}


In this work, we perform all extrapolations by using a non-linear
least-squares fit to a form that is specific to each observable and
varies with cutoff parameter. The fitting process is iterated until
all fit parameters have converged to at least 10 digits of precision.
We apply forms identically for both NCSM and MR-IMSRG extrapolations,
treating the former as functions of $N_{\text{max}}$ and the latter
as functions of $e_{\text{max}}$. We use $X_{\text{max}}$ to denote
either cutoff parameter when defining the extrapolations provided
below. Following a common NCFC practice, we do not include the $N_{\text{max}}=0$
result when performing fits to any of the NCSM data sets. The extrapolation
for each data set is performed without regard to any other data sets
or their extrapolations. The formulae for ground-state energy and
square radius are applied identically to both NCSM and MR-IMSRG results.
Extrapolations for $0\nu\beta\beta$-decay NMEs are only performed
for the NCSM results because of fluctuations in the MR-IMSRG results.

It should be noted that the extrapolations described here were originally
designed with the $N_{\text{max}}$ truncation scheme in mind, and
their effectiveness for extrapolating results in the $e_{\text{max}}$
truncation scheme has not yet been fully explored. Nevertheless, the
significant similarities of the two schemes and their quantification
of the same underlying variable (i.e. the content of the many-body
basis) suggest the same extrapolation forms may be effective; an expectation
that is supported by the results of this work.

Motivated by the extrapolations proposed in Ref.$\ $\citealp{Maris:2008ax},
we extrapolate the ground-state energy to the form 
\begin{flalign}
f\left(X_{\text{max}}\right) & =a+b\cdot e^{-cX_{\text{max}}}\label{eq:0nbb ext.}
\end{flalign}
where $a$, $b$, and $c$ are fit parameters. We employ the same
form for our extrapolations of the NCSM $0\nu\beta\beta$-decay results.
Values of fit parameters calculated in this study for energy and $0\nu\beta\beta$-decay
NMEs may be found in the right-most columns of Table$\ $(\ref{table:1})
alongside their corresponding data set.

The simple exponential form depicted in Eq.$\ $(\ref{eq:0nbb ext.})
generally provides a poor prediction for the convergence behavior
of square-radius operator observables. Thus, inspired by the methods
discussed in Ref.$\ $\citealp{Furnstahl:2012qg}, we extrapolate
square radii by fitting to the form 
\begin{flalign}
\langle r^{2}\rangle & =\langle r^{2}\rangle_{\infty}-\left(c_{0}\beta+c_{1}\beta^{3}\right)e^{-\beta}\,,\label{eq:rms R ext.}
\end{flalign}
where {\small{}
\begin{flalign*}
\beta\equiv & 2k_{\infty}\frac{\hbar}{m\Omega}\left[\sqrt{2X_{\text{max}}+5}+0.54437\left(2X_{\text{max}}+5\right)^{1/6}\right]\,.
\end{flalign*}
}Here $m=938.92$ MeV is the average mass of a neutron and a proton,
and $\langle r^{2}\rangle_{\infty},$ $c_{0}$, and $c_{1}$ are fit
parameters. Unlike the authors of Ref.$\ $\citealp{Furnstahl:2012qg}
who determine $k_{\infty}$ while extrapolating the ground-state energy
with their theoretically-founded ``IR formula'', we treat $k_{\infty}$
as an additional fit parameter when extrapolating each square radius.
We provide our calculated values of the fit parameters for each square
radius extrapolation alongside its corresponding data set in Table$\ $(\ref{table:2}).
\begin{table*}[b]
\caption{\label{table:1}MR-IMSRG(2) and NCSM calculated observables and extrapolation
parameters (see Eq.$\ $(\ref{eq:0nbb ext.})). The ground-state energy
($E$) results correspond to the $^{6}\text{He}$ ground state, and
are plotted in Fig.$\ $(\ref{fig:1}). The $0\nu\beta\beta$-decay
NME ($M_{0\nu}$) results and their decomposition into Fermi (F),
Gamow-Teller (GT), and tensor (T) contributions correspond to the$^{6}\text{He}\rightarrow{}^{6}\text{Be}$
ground-state-to-ground-state transition, and are plotted in Fig.$\ $(\ref{fig:2}).
In all calculations of $M_{0\nu}$ and its contributions, isospin
symmetry has been assumed. Extrapolations for the $0\nu\beta\beta$-decay
NME were only performed using the NCSM results.}
\begin{ruledtabular}
\begin{tabular}{l|l||ccccccc|ccc}
\multicolumn{1}{l}{} &  & \multicolumn{7}{c|}{$N_{\text{max}}$($e_{\text{max}}$)} & \multicolumn{3}{c}{Fit Parameters}\tabularnewline
\hline 
Observable & Method & 0(2) & 2(4) & 4(6) & 6(8) & 8(10) & 10(12) & 12 & $a$ & $b$ & $c$\tabularnewline
\hline 
\multirow{2}{*}{$E$ (MeV) } & NCSM & -12.546 & -19.406 & -23.961 & -26.438 & -27.699 & -28.374 & -28.720 & -29.132 & 18.414 & 0.3188\tabularnewline
 & MR-IMSRG & -20.810 & -26.037 & -27.752 & -28.240 & -28.385 & -28.435 &  & -28.472 & 24.375 & 0.5784\tabularnewline
\hline 
\multirow{2}{*}{$M_{0\nu}^{F}$} & NCSM & -1.0165 & -0.9669 & -0.9287 & -0.8984 & -0.8773 & -0.8604 & -0.8458 & -0.8032 & -0.2135 & 0.1331\tabularnewline
 & MR-IMSRG & -1.0430 & -0.9811 & -0.9335 & -0.9452 & -0.8880 & -0.9110 &  &  &  & \tabularnewline
\hline 
\multirow{2}{*}{$M_{0\nu}^{GT}$} & NCSM & -4.0553 & -3.7751 & -3.6398 & -3.5256 & -3.4546 & -3.3960 & -3.3471 & -3.2144 & -0.7472 & 0.1429\tabularnewline
 & MR-IMSRG & -3.9576 & -3.7688 & -3.5812 & -3.7742 & -3.4503 & -3.5326 &  &  &  & \tabularnewline
\hline 
\multirow{2}{*}{$M_{0\nu}^{T}$} & NCSM & 0.0435 & 0.0479 & 0.0325 & 0.0313 & 0.0279 & 0.0265 & 0.0252 & 0.0247 & 0.0441 & 0.3472\tabularnewline
 & MR-IMSRG & 0.0661 & 0.0484 & 0.0441 & 0.0268 & 0.0294 & 0.0301 &  &  &  & \tabularnewline
\hline 
\multirow{2}{*}{$M_{0\nu}$} & NCSM & -5.0283 & -4.6941 & -4.5361 & -4.3926 & -4.3041 & -4.2299 & -4.1677 & -3.9777 & -0.9355 & 0.1322\tabularnewline
 & MR-IMSRG & -4.9346 & -4.7016 & -4.4706 & -4.6927 & -4.3089 & -4.4134 &  &  &  & \tabularnewline
\end{tabular}
\end{ruledtabular}

\end{table*}
\begin{table*}[b]
\caption{\label{table:2}MR-IMSRG(2) and NCSM neutron, proton, and matter square
radii ($r_{n}^{2}$, $r_{p}^{2}$, and $r_{m}^{2}$, respectively)
and corresponding square radius extrapolation parameters (see Eq.$\ $(\ref{eq:rms R ext.}))
for the $^{6}\text{He}$ ground state. The extrapolated fits are plotted
alongside their respective results in Fig.$\ $(\ref{fig:1}).}
\begin{ruledtabular}
\begin{tabular}{l|l||ccccccc|cccc}
\multicolumn{1}{l}{} &  & \multicolumn{7}{c|}{$N_{\text{max}}$($e_{\text{max}}$)} & \multicolumn{4}{c}{Fit Parameters }\tabularnewline
\hline 
Observable & Method & 0(2) & 2(4) & 4(6) & 6(8) & 8(10) & 10(12) & 12 & $\langle r^{2}\rangle_{\infty}$ & $c_{0}$ & $c_{1}$ & $k_{\infty}$\tabularnewline
\hline 
\multirow{2}{*}{$r_{n}^{2}$ ($\text{fm}^{2}$)} & NCSM & 3.6286 & 4.0096 & 4.3347 & 4.6604 & 4.9169 & 5.1656 & 5.4014 & 8.0927 & 10.9067 & 0.9965 & 0.1238\tabularnewline
 & MR-IMSRG & 3.8389 & 4.3023 & 4.5616 & 4.7524 & 4.8929 & 4.9903 &  & 5.3226 & 10.601 & 0.2827 & 0.2235\tabularnewline
\hline 
\multirow{2}{*}{$r_{p}^{2}$ ($\text{fm}^{2}$)} & NCSM & 2.5918 & 2.5622 & 2.6955 & 2.8056 & 2.8791 & 2.9532 & 3.0217 & 3.4732 & 3.0146 & 0.1924 & 0.1505\tabularnewline
 & MR-IMSRG & 2.5125 & 2.6540 & 2.6870 & 2.7149 & 2.7394 & 2.7543 &  & 2.7598 & 8.9888 & 0.0807 & 0.3546\tabularnewline
\hline 
\multirow{2}{*}{$r_{m}^{2}$ ($\text{fm}^{2}$)} & NCSM & 3.2830 & 3.5273 & 3.7885 & 4.0421 & 4.2378 & 4.4285 & 4.6083 & 6.5005 & 8.1034 & 0.7285 & 0.1282\tabularnewline
 & MR-IMSRG & 3.3970 & 3.7527 & 3.9367 & 4.0733 & 4.1751 & 4.2448 &  & 4.4563 & 8.4272 & 0.2028 & 0.2326\tabularnewline
\end{tabular}
\end{ruledtabular}

\end{table*}

\bibliographystyle{apsrev4-1}

\bibliography{references}

\providecommand{\noopsort}[1]{}\providecommand{\singleletter}[1]{#1}
\begin{thebibliography}{76}%
\makeatletter
\providecommand \@ifxundefined [1]{%
 \@ifx{#1\undefined}
}%
\providecommand \@ifnum [1]{%
 \ifnum #1\expandafter \@firstoftwo
 \else \expandafter \@secondoftwo
 \fi
}%
\providecommand \@ifx [1]{%
 \ifx #1\expandafter \@firstoftwo
 \else \expandafter \@secondoftwo
 \fi
}%
\providecommand \natexlab [1]{#1}%
\providecommand \enquote  [1]{``#1''}%
\providecommand \bibnamefont  [1]{#1}%
\providecommand \bibfnamefont [1]{#1}%
\providecommand \citenamefont [1]{#1}%
\providecommand \href@noop [0]{\@secondoftwo}%
\providecommand \href [0]{\begingroup \@sanitize@url \@href}%
\providecommand \@href[1]{\@@startlink{#1}\@@href}%
\providecommand \@@href[1]{\endgroup#1\@@endlink}%
\providecommand \@sanitize@url [0]{\catcode `\\12\catcode `\$12\catcode
  `\&12\catcode `\#12\catcode `\^12\catcode `\_12\catcode `\%12\relax}%
\providecommand \@@startlink[1]{}%
\providecommand \@@endlink[0]{}%
\providecommand \url  [0]{\begingroup\@sanitize@url \@url }%
\providecommand \@url [1]{\endgroup\@href {#1}{\urlprefix }}%
\providecommand \urlprefix  [0]{URL }%
\providecommand \Eprint [0]{\href }%
\providecommand \doibase [0]{http://dx.doi.org/}%
\providecommand \selectlanguage [0]{\@gobble}%
\providecommand \bibinfo  [0]{\@secondoftwo}%
\providecommand \bibfield  [0]{\@secondoftwo}%
\providecommand \translation [1]{[#1]}%
\providecommand \BibitemOpen [0]{}%
\providecommand \bibitemStop [0]{}%
\providecommand \bibitemNoStop [0]{.\EOS\space}%
\providecommand \EOS [0]{\spacefactor3000\relax}%
\providecommand \BibitemShut  [1]{\csname bibitem#1\endcsname}%
\let\auto@bib@innerbib\@empty
\bibitem [{\citenamefont {Ahmad}\ \emph {et~al.}(2001)\citenamefont {Ahmad}
  \emph {et~al.}}]{Ahmad:2001an}%
  \BibitemOpen
  \bibfield  {author} {\bibinfo {author} {\bibfnamefont {Q.~R.}\ \bibnamefont
  {Ahmad}} \emph {et~al.} (\bibinfo {collaboration} {SNO}),\ }\href {\doibase
  10.1103/PhysRevLett.87.071301} {\bibfield  {journal} {\bibinfo  {journal}
  {Phys. Rev. Lett.}\ }\textbf {\bibinfo {volume} {87}},\ \bibinfo {pages}
  {071301} (\bibinfo {year} {2001})},\ \Eprint
  {http://arxiv.org/abs/nucl-ex/0106015} {arXiv:nucl-ex/0106015 [nucl-ex]}
  \BibitemShut {NoStop}%
\bibitem [{\citenamefont {Eguchi}\ \emph {et~al.}(2003)\citenamefont {Eguchi}
  \emph {et~al.}}]{Eguchi:2002dm}%
  \BibitemOpen
  \bibfield  {author} {\bibinfo {author} {\bibfnamefont {K.}~\bibnamefont
  {Eguchi}} \emph {et~al.} (\bibinfo {collaboration} {KamLAND}),\ }\href
  {\doibase 10.1103/PhysRevLett.90.021802} {\bibfield  {journal} {\bibinfo
  {journal} {Phys. Rev. Lett.}\ }\textbf {\bibinfo {volume} {90}},\ \bibinfo
  {pages} {021802} (\bibinfo {year} {2003})},\ \Eprint
  {http://arxiv.org/abs/hep-ex/0212021} {arXiv:hep-ex/0212021 [hep-ex]}
  \BibitemShut {NoStop}%
\bibitem [{\citenamefont {Fukuda}\ \emph {et~al.}(1998)\citenamefont {Fukuda}
  \emph {et~al.}}]{Fukuda:1998mi}%
  \BibitemOpen
  \bibfield  {author} {\bibinfo {author} {\bibfnamefont {Y.}~\bibnamefont
  {Fukuda}} \emph {et~al.} (\bibinfo {collaboration} {Super-Kamiokande}),\
  }\href {\doibase 10.1103/PhysRevLett.81.1562} {\bibfield  {journal} {\bibinfo
   {journal} {Phys. Rev. Lett.}\ }\textbf {\bibinfo {volume} {81}},\ \bibinfo
  {pages} {1562} (\bibinfo {year} {1998})},\ \Eprint
  {http://arxiv.org/abs/hep-ex/9807003} {arXiv:hep-ex/9807003 [hep-ex]}
  \BibitemShut {NoStop}%
\bibitem [{\citenamefont {Martin-Albo}\ \emph {et~al.}(2016)\citenamefont
  {Martin-Albo} \emph {et~al.}}]{Martin-Albo:2015rhw}%
  \BibitemOpen
  \bibfield  {author} {\bibinfo {author} {\bibfnamefont {J.}~\bibnamefont
  {Martin-Albo}} \emph {et~al.} (\bibinfo {collaboration} {NEXT}),\ }\href
  {\doibase 10.1007/JHEP05(2016)159} {\bibfield  {journal} {\bibinfo  {journal}
  {JHEP}\ }\textbf {\bibinfo {volume} {05}},\ \bibinfo {pages} {159} (\bibinfo
  {year} {2016})},\ \Eprint {http://arxiv.org/abs/1511.09246} {arXiv:1511.09246
  [physics.ins-det]} \BibitemShut {NoStop}%
\bibitem [{\citenamefont {Albert}\ \emph {et~al.}(2014)\citenamefont {Albert}
  \emph {et~al.}}]{Albert:2014awa}%
  \BibitemOpen
  \bibfield  {author} {\bibinfo {author} {\bibfnamefont {J.~B.}\ \bibnamefont
  {Albert}} \emph {et~al.} (\bibinfo {collaboration} {EXO-200}),\ }\href
  {\doibase 10.1038/nature13432} {\bibfield  {journal} {\bibinfo  {journal}
  {Nature}\ }\textbf {\bibinfo {volume} {510}},\ \bibinfo {pages} {229}
  (\bibinfo {year} {2014})},\ \Eprint {http://arxiv.org/abs/1402.6956}
  {arXiv:1402.6956 [nucl-ex]} \BibitemShut {NoStop}%
\bibitem [{\citenamefont {Gilliss}\ \emph {et~al.}(2018)\citenamefont {Gilliss}
  \emph {et~al.}}]{Gilliss:2018lke}%
  \BibitemOpen
  \bibfield  {author} {\bibinfo {author} {\bibfnamefont {T.}~\bibnamefont
  {Gilliss}} \emph {et~al.} (\bibinfo {collaboration} {MAJORANA}),\ }\bibfield
  {booktitle} {\emph {\bibinfo {booktitle} {{Proceedings, 21st International
  Conference on Particles and Nuclei (PANIC 17): Beijing, China, September 1-5,
  2017}}},\ }\href {\doibase 10.1142/S2010194518600492} {\bibfield  {journal}
  {\bibinfo  {journal} {Int. J. Mod. Phys. Conf. Ser.}\ }\textbf {\bibinfo
  {volume} {46}},\ \bibinfo {pages} {1860049} (\bibinfo {year} {2018})},\
  \Eprint {http://arxiv.org/abs/1804.01582} {arXiv:1804.01582
  [physics.ins-det]} \BibitemShut {NoStop}%
\bibitem [{\citenamefont {Gando}\ \emph {et~al.}(2016)\citenamefont {Gando},
  \citenamefont {Gando}, \citenamefont {Hachiya}, \citenamefont {Hayashi},
  \citenamefont {Hayashida}, \citenamefont {Ikeda}, \citenamefont {Inoue},
  \citenamefont {Ishidoshiro}, \citenamefont {Karino} \emph
  {et~al.}}]{KamLAND-Zen:2016pfg}%
  \BibitemOpen
  \bibfield  {author} {\bibinfo {author} {\bibfnamefont {A.}~\bibnamefont
  {Gando}}, \bibinfo {author} {\bibfnamefont {Y.}~\bibnamefont {Gando}},
  \bibinfo {author} {\bibfnamefont {T.}~\bibnamefont {Hachiya}}, \bibinfo
  {author} {\bibfnamefont {A.}~\bibnamefont {Hayashi}}, \bibinfo {author}
  {\bibfnamefont {S.}~\bibnamefont {Hayashida}}, \bibinfo {author}
  {\bibfnamefont {H.}~\bibnamefont {Ikeda}}, \bibinfo {author} {\bibfnamefont
  {K.}~\bibnamefont {Inoue}}, \bibinfo {author} {\bibfnamefont
  {K.}~\bibnamefont {Ishidoshiro}}, \bibinfo {author} {\bibfnamefont
  {Y.}~\bibnamefont {Karino}},  \emph {et~al.} (\bibinfo {collaboration}
  {KamLAND-Zen}),\ }\href {\doibase 10.1103/PhysRevLett.117.109903,
  10.1103/PhysRevLett.117.082503} {\bibfield  {journal} {\bibinfo  {journal}
  {Phys. Rev. Lett.}\ }\textbf {\bibinfo {volume} {117}},\ \bibinfo {pages}
  {082503} (\bibinfo {year} {2016})},\ \bibinfo {note} {[Addendum: Phys. Rev.
  Lett.117,no.10,109903(2016)]},\ \Eprint {http://arxiv.org/abs/1605.02889}
  {arXiv:1605.02889 [hep-ex]} \BibitemShut {NoStop}%
\bibitem [{\citenamefont {Alfonso}\ \emph {et~al.}(2015)\citenamefont {Alfonso}
  \emph {et~al.}}]{Alfonso:2015wka}%
  \BibitemOpen
  \bibfield  {author} {\bibinfo {author} {\bibfnamefont {K.}~\bibnamefont
  {Alfonso}} \emph {et~al.} (\bibinfo {collaboration} {CUORE}),\ }\href
  {\doibase 10.1103/PhysRevLett.115.102502} {\bibfield  {journal} {\bibinfo
  {journal} {Phys. Rev. Lett.}\ }\textbf {\bibinfo {volume} {115}},\ \bibinfo
  {pages} {102502} (\bibinfo {year} {2015})},\ \Eprint
  {http://arxiv.org/abs/1504.02454} {arXiv:1504.02454 [nucl-ex]} \BibitemShut
  {NoStop}%
\bibitem [{\citenamefont {Agostini}\ \emph {et~al.}(2017)\citenamefont
  {Agostini} \emph {et~al.}}]{Agostini:2017iyd}%
  \BibitemOpen
  \bibfield  {author} {\bibinfo {author} {\bibfnamefont {M.}~\bibnamefont
  {Agostini}} \emph {et~al.},\ }\href {\doibase 10.1038/nature21717} {\
  (\bibinfo {year} {2017}),\ 10.1038/nature21717},\ \bibinfo {note}
  {[Nature544,47(2017)]},\ \Eprint {http://arxiv.org/abs/1703.00570}
  {arXiv:1703.00570 [nucl-ex]} \BibitemShut {NoStop}%
\bibitem [{\citenamefont {Aalseth}\ \emph {et~al.}(2018)\citenamefont {Aalseth}
  \emph {et~al.}}]{Aalseth:2017btx}%
  \BibitemOpen
  \bibfield  {author} {\bibinfo {author} {\bibfnamefont {C.~E.}\ \bibnamefont
  {Aalseth}} \emph {et~al.} (\bibinfo {collaboration} {Majorana}),\ }\href
  {\doibase 10.1103/PhysRevLett.120.132502} {\bibfield  {journal} {\bibinfo
  {journal} {Phys. Rev. Lett.}\ }\textbf {\bibinfo {volume} {120}},\ \bibinfo
  {pages} {132502} (\bibinfo {year} {2018})},\ \Eprint
  {http://arxiv.org/abs/1710.11608} {arXiv:1710.11608 [nucl-ex]} \BibitemShut
  {NoStop}%
\bibitem [{\citenamefont {Andringa}\ \emph {et~al.}(2016)\citenamefont
  {Andringa} \emph {et~al.}}]{Andringa:2015tza}%
  \BibitemOpen
  \bibfield  {author} {\bibinfo {author} {\bibfnamefont {S.}~\bibnamefont
  {Andringa}} \emph {et~al.} (\bibinfo {collaboration} {SNO+}),\ }\href
  {\doibase 10.1155/2016/6194250} {\bibfield  {journal} {\bibinfo  {journal}
  {Adv. High Energy Phys.}\ }\textbf {\bibinfo {volume} {2016}},\ \bibinfo
  {pages} {6194250} (\bibinfo {year} {2016})},\ \Eprint
  {http://arxiv.org/abs/1508.05759} {arXiv:1508.05759 [physics.ins-det]}
  \BibitemShut {NoStop}%
\bibitem [{\citenamefont {Iwata}\ \emph {et~al.}(2016)\citenamefont {Iwata},
  \citenamefont {Shimizu}, \citenamefont {Otsuka}, \citenamefont {Utsuno},
  \citenamefont {Men{\`e}ndez}, \citenamefont {Honma},\ and\ \citenamefont
  {Abe}}]{Iwata:2016cxn}%
  \BibitemOpen
  \bibfield  {author} {\bibinfo {author} {\bibfnamefont {Y.}~\bibnamefont
  {Iwata}}, \bibinfo {author} {\bibfnamefont {N.}~\bibnamefont {Shimizu}},
  \bibinfo {author} {\bibfnamefont {T.}~\bibnamefont {Otsuka}}, \bibinfo
  {author} {\bibfnamefont {Y.}~\bibnamefont {Utsuno}}, \bibinfo {author}
  {\bibfnamefont {J.}~\bibnamefont {Men{\`e}ndez}}, \bibinfo {author}
  {\bibfnamefont {M.}~\bibnamefont {Honma}}, \ and\ \bibinfo {author}
  {\bibfnamefont {T.}~\bibnamefont {Abe}},\ }\href {\doibase
  10.1103/PhysRevLett.117.179902, 10.1103/PhysRevLett.116.112502} {\bibfield
  {journal} {\bibinfo  {journal} {Phys. Rev. Lett.}\ }\textbf {\bibinfo
  {volume} {116}},\ \bibinfo {pages} {112502} (\bibinfo {year} {2016})},\
  \bibinfo {note} {[Erratum: Phys. Rev. Lett.117,no.17,179902(2016)]},\ \Eprint
  {http://arxiv.org/abs/1602.07822} {arXiv:1602.07822 [nucl-th]} \BibitemShut
  {NoStop}%
\bibitem [{\citenamefont {Cirigliano}\ \emph {et~al.}(2017)\citenamefont
  {Cirigliano}, \citenamefont {Dekens}, \citenamefont {de~Vries}, \citenamefont
  {Graesser},\ and\ \citenamefont {Mereghetti}}]{Cirigliano:2017djv}%
  \BibitemOpen
  \bibfield  {author} {\bibinfo {author} {\bibfnamefont {V.}~\bibnamefont
  {Cirigliano}}, \bibinfo {author} {\bibfnamefont {W.}~\bibnamefont {Dekens}},
  \bibinfo {author} {\bibfnamefont {J.}~\bibnamefont {de~Vries}}, \bibinfo
  {author} {\bibfnamefont {M.~L.}\ \bibnamefont {Graesser}}, \ and\ \bibinfo
  {author} {\bibfnamefont {E.}~\bibnamefont {Mereghetti}},\ }\href {\doibase
  10.1007/JHEP12(2017)082} {\bibfield  {journal} {\bibinfo  {journal} {JHEP}\
  }\textbf {\bibinfo {volume} {12}},\ \bibinfo {pages} {082} (\bibinfo {year}
  {2017})},\ \Eprint {http://arxiv.org/abs/1708.09390} {arXiv:1708.09390
  [hep-ph]} \BibitemShut {NoStop}%
\bibitem [{\citenamefont {Contessi}\ \emph {et~al.}(2017)\citenamefont
  {Contessi}, \citenamefont {Lovato}, \citenamefont {Pederiva}, \citenamefont
  {Roggero}, \citenamefont {Kirscher},\ and\ \citenamefont {van
  Kolck}}]{Contessi:2017rww}%
  \BibitemOpen
  \bibfield  {author} {\bibinfo {author} {\bibfnamefont {L.}~\bibnamefont
  {Contessi}}, \bibinfo {author} {\bibfnamefont {A.}~\bibnamefont {Lovato}},
  \bibinfo {author} {\bibfnamefont {F.}~\bibnamefont {Pederiva}}, \bibinfo
  {author} {\bibfnamefont {A.}~\bibnamefont {Roggero}}, \bibinfo {author}
  {\bibfnamefont {J.}~\bibnamefont {Kirscher}}, \ and\ \bibinfo {author}
  {\bibfnamefont {U.}~\bibnamefont {van Kolck}},\ }\href {\doibase
  10.1016/j.physletb.2017.07.048} {\bibfield  {journal} {\bibinfo  {journal}
  {Phys. Lett.}\ }\textbf {\bibinfo {volume} {B772}},\ \bibinfo {pages} {839}
  (\bibinfo {year} {2017})},\ \Eprint {http://arxiv.org/abs/1701.06516}
  {arXiv:1701.06516 [nucl-th]} \BibitemShut {NoStop}%
\bibitem [{\citenamefont {Coraggio}\ \emph {et~al.}(2017)\citenamefont
  {Coraggio}, \citenamefont {De~Angelis}, \citenamefont {Fukui}, \citenamefont
  {Gargano},\ and\ \citenamefont {Itaco}}]{Coraggio:2017bqn}%
  \BibitemOpen
  \bibfield  {author} {\bibinfo {author} {\bibfnamefont {L.}~\bibnamefont
  {Coraggio}}, \bibinfo {author} {\bibfnamefont {L.}~\bibnamefont
  {De~Angelis}}, \bibinfo {author} {\bibfnamefont {T.}~\bibnamefont {Fukui}},
  \bibinfo {author} {\bibfnamefont {A.}~\bibnamefont {Gargano}}, \ and\
  \bibinfo {author} {\bibfnamefont {N.}~\bibnamefont {Itaco}},\ }\href
  {\doibase 10.1103/PhysRevC.95.064324} {\bibfield  {journal} {\bibinfo
  {journal} {Phys. Rev.}\ }\textbf {\bibinfo {volume} {C95}},\ \bibinfo {pages}
  {064324} (\bibinfo {year} {2017})},\ \Eprint
  {http://arxiv.org/abs/1703.05087} {arXiv:1703.05087 [nucl-th]} \BibitemShut
  {NoStop}%
\bibitem [{\citenamefont {Jiao}\ \emph {et~al.}(2017)\citenamefont {Jiao},
  \citenamefont {Engel},\ and\ \citenamefont {Holt}}]{Jiao:2017opc}%
  \BibitemOpen
  \bibfield  {author} {\bibinfo {author} {\bibfnamefont {C.~F.}\ \bibnamefont
  {Jiao}}, \bibinfo {author} {\bibfnamefont {J.}~\bibnamefont {Engel}}, \ and\
  \bibinfo {author} {\bibfnamefont {J.~D.}\ \bibnamefont {Holt}},\ }\href
  {\doibase 10.1103/PhysRevC.96.054310} {\bibfield  {journal} {\bibinfo
  {journal} {Phys. Rev.}\ }\textbf {\bibinfo {volume} {C96}},\ \bibinfo {pages}
  {054310} (\bibinfo {year} {2017})},\ \Eprint
  {http://arxiv.org/abs/1707.03940} {arXiv:1707.03940 [nucl-th]} \BibitemShut
  {NoStop}%
\bibitem [{\citenamefont {Tiburzi}\ \emph {et~al.}(2017)\citenamefont
  {Tiburzi}, \citenamefont {Wagman}, \citenamefont {Winter}, \citenamefont
  {Chang}, \citenamefont {Davoudi}, \citenamefont {Detmold}, \citenamefont
  {Orginos}, \citenamefont {Savage},\ and\ \citenamefont
  {Shanahan}}]{Tiburzi:2017iux}%
  \BibitemOpen
  \bibfield  {author} {\bibinfo {author} {\bibfnamefont {B.~C.}\ \bibnamefont
  {Tiburzi}}, \bibinfo {author} {\bibfnamefont {M.~L.}\ \bibnamefont {Wagman}},
  \bibinfo {author} {\bibfnamefont {F.}~\bibnamefont {Winter}}, \bibinfo
  {author} {\bibfnamefont {E.}~\bibnamefont {Chang}}, \bibinfo {author}
  {\bibfnamefont {Z.}~\bibnamefont {Davoudi}}, \bibinfo {author} {\bibfnamefont
  {W.}~\bibnamefont {Detmold}}, \bibinfo {author} {\bibfnamefont
  {K.}~\bibnamefont {Orginos}}, \bibinfo {author} {\bibfnamefont {M.~J.}\
  \bibnamefont {Savage}}, \ and\ \bibinfo {author} {\bibfnamefont {P.~E.}\
  \bibnamefont {Shanahan}},\ }\href {\doibase 10.1103/PhysRevD.96.054505}
  {\bibfield  {journal} {\bibinfo  {journal} {Phys. Rev.}\ }\textbf {\bibinfo
  {volume} {D96}},\ \bibinfo {pages} {054505} (\bibinfo {year} {2017})},\
  \Eprint {http://arxiv.org/abs/1702.02929} {arXiv:1702.02929 [hep-lat]}
  \BibitemShut {NoStop}%
\bibitem [{\citenamefont {Shanahan}\ \emph {et~al.}(2017)\citenamefont
  {Shanahan}, \citenamefont {Tiburzi}, \citenamefont {Wagman}, \citenamefont
  {Winter}, \citenamefont {Chang}, \citenamefont {Davoudi}, \citenamefont
  {Detmold}, \citenamefont {Orginos},\ and\ \citenamefont
  {Savage}}]{Shanahan:2017bgi}%
  \BibitemOpen
  \bibfield  {author} {\bibinfo {author} {\bibfnamefont {P.~E.}\ \bibnamefont
  {Shanahan}}, \bibinfo {author} {\bibfnamefont {B.~C.}\ \bibnamefont
  {Tiburzi}}, \bibinfo {author} {\bibfnamefont {M.~L.}\ \bibnamefont {Wagman}},
  \bibinfo {author} {\bibfnamefont {F.}~\bibnamefont {Winter}}, \bibinfo
  {author} {\bibfnamefont {E.}~\bibnamefont {Chang}}, \bibinfo {author}
  {\bibfnamefont {Z.}~\bibnamefont {Davoudi}}, \bibinfo {author} {\bibfnamefont
  {W.}~\bibnamefont {Detmold}}, \bibinfo {author} {\bibfnamefont
  {K.}~\bibnamefont {Orginos}}, \ and\ \bibinfo {author} {\bibfnamefont
  {M.~J.}\ \bibnamefont {Savage}},\ }\href {\doibase
  10.1103/PhysRevLett.119.062003} {\bibfield  {journal} {\bibinfo  {journal}
  {Phys. Rev. Lett.}\ }\textbf {\bibinfo {volume} {119}},\ \bibinfo {pages}
  {062003} (\bibinfo {year} {2017})},\ \Eprint
  {http://arxiv.org/abs/1701.03456} {arXiv:1701.03456 [hep-lat]} \BibitemShut
  {NoStop}%
\bibitem [{\citenamefont {Horoi}\ and\ \citenamefont
  {Neacsu}(2017)}]{Horoi:2017gmj}%
  \BibitemOpen
  \bibfield  {author} {\bibinfo {author} {\bibfnamefont {M.}~\bibnamefont
  {Horoi}}\ and\ \bibinfo {author} {\bibfnamefont {A.}~\bibnamefont {Neacsu}},\
  }\href@noop {} {\  (\bibinfo {year} {2017})},\ \Eprint
  {http://arxiv.org/abs/1706.05391} {arXiv:1706.05391 [hep-ph]} \BibitemShut
  {NoStop}%
\bibitem [{\citenamefont {Cremonesi}\ and\ \citenamefont
  {Pavan}(2014)}]{Cremonesi:2013vla}%
  \BibitemOpen
  \bibfield  {author} {\bibinfo {author} {\bibfnamefont {O.}~\bibnamefont
  {Cremonesi}}\ and\ \bibinfo {author} {\bibfnamefont {M.}~\bibnamefont
  {Pavan}},\ }\href {\doibase 10.1155/2014/951432} {\bibfield  {journal}
  {\bibinfo  {journal} {Adv. High Energy Phys.}\ }\textbf {\bibinfo {volume}
  {2014}},\ \bibinfo {pages} {951432} (\bibinfo {year} {2014})},\ \Eprint
  {http://arxiv.org/abs/1310.4692} {arXiv:1310.4692 [physics.ins-det]}
  \BibitemShut {NoStop}%
\bibitem [{\citenamefont {Dell'Oro}\ \emph {et~al.}(2016)\citenamefont
  {Dell'Oro}, \citenamefont {Marcocci}, \citenamefont {Viel},\ and\
  \citenamefont {Vissani}}]{DellOro:2016tmg}%
  \BibitemOpen
  \bibfield  {author} {\bibinfo {author} {\bibfnamefont {S.}~\bibnamefont
  {Dell'Oro}}, \bibinfo {author} {\bibfnamefont {S.}~\bibnamefont {Marcocci}},
  \bibinfo {author} {\bibfnamefont {M.}~\bibnamefont {Viel}}, \ and\ \bibinfo
  {author} {\bibfnamefont {F.}~\bibnamefont {Vissani}},\ }\href {\doibase
  10.1155/2016/2162659} {\bibfield  {journal} {\bibinfo  {journal} {Adv. High
  Energy Phys.}\ }\textbf {\bibinfo {volume} {2016}},\ \bibinfo {pages}
  {2162659} (\bibinfo {year} {2016})},\ \Eprint
  {http://arxiv.org/abs/1601.07512} {arXiv:1601.07512 [hep-ph]} \BibitemShut
  {NoStop}%
\bibitem [{\citenamefont {Engel}\ and\ \citenamefont
  {Men{\`e}ndez}(2017)}]{Engel:2016xgb}%
  \BibitemOpen
  \bibfield  {author} {\bibinfo {author} {\bibfnamefont {J.}~\bibnamefont
  {Engel}}\ and\ \bibinfo {author} {\bibfnamefont {J.}~\bibnamefont
  {Men{\`e}ndez}},\ }\href {\doibase 10.1088/1361-6633/aa5bc5} {\bibfield
  {journal} {\bibinfo  {journal} {Rept. Prog. Phys.}\ }\textbf {\bibinfo
  {volume} {80}},\ \bibinfo {pages} {046301} (\bibinfo {year} {2017})},\
  \Eprint {http://arxiv.org/abs/1610.06548} {arXiv:1610.06548 [nucl-th]}
  \BibitemShut {NoStop}%
\bibitem [{\citenamefont {Gomez-Cadenas}\ \emph {et~al.}(2012)\citenamefont
  {Gomez-Cadenas}, \citenamefont {Martin-Albo}, \citenamefont {Mezzetto},
  \citenamefont {Monrabal},\ and\ \citenamefont {Sorel}}]{GomezCadenas:2011it}%
  \BibitemOpen
  \bibfield  {author} {\bibinfo {author} {\bibfnamefont {J.~J.}\ \bibnamefont
  {Gomez-Cadenas}}, \bibinfo {author} {\bibfnamefont {J.}~\bibnamefont
  {Martin-Albo}}, \bibinfo {author} {\bibfnamefont {M.}~\bibnamefont
  {Mezzetto}}, \bibinfo {author} {\bibfnamefont {F.}~\bibnamefont {Monrabal}},
  \ and\ \bibinfo {author} {\bibfnamefont {M.}~\bibnamefont {Sorel}},\ }\href
  {\doibase 10.1393/ncr/i2012-10074-9} {\bibfield  {journal} {\bibinfo
  {journal} {Riv. Nuovo Cim.}\ }\textbf {\bibinfo {volume} {35}},\ \bibinfo
  {pages} {29} (\bibinfo {year} {2012})},\ \Eprint
  {http://arxiv.org/abs/1109.5515} {arXiv:1109.5515 [hep-ex]} \BibitemShut
  {NoStop}%
\bibitem [{\citenamefont {Henning}(2016)}]{Henning:2016fad}%
  \BibitemOpen
  \bibfield  {author} {\bibinfo {author} {\bibfnamefont {R.}~\bibnamefont
  {Henning}},\ }\href {\doibase 10.1016/j.revip.2016.03.001} {\bibfield
  {journal} {\bibinfo  {journal} {Rev. Phys.}\ }\textbf {\bibinfo {volume}
  {1}},\ \bibinfo {pages} {29} (\bibinfo {year} {2016})}\BibitemShut {NoStop}%
\bibitem [{\citenamefont {Gysbers}\ \emph {et~al.}(2019)\citenamefont {Gysbers}
  \emph {et~al.}}]{Gysbers:2019uyb}%
  \BibitemOpen
  \bibfield  {author} {\bibinfo {author} {\bibfnamefont {P.}~\bibnamefont
  {Gysbers}} \emph {et~al.},\ }\href {\doibase 10.1038/s41567-019-0450-7}
  {\bibfield  {journal} {\bibinfo  {journal} {Nature Phys.}\ }\textbf {\bibinfo
  {volume} {15}},\ \bibinfo {pages} {428} (\bibinfo {year} {2019})},\ \Eprint
  {http://arxiv.org/abs/1903.00047} {arXiv:1903.00047 [nucl-th]} \BibitemShut
  {NoStop}%
\bibitem [{\citenamefont {Bogner}\ \emph {et~al.}(2014)\citenamefont {Bogner},
  \citenamefont {Hergert}, \citenamefont {Holt}, \citenamefont {Schwenk},
  \citenamefont {Binder}, \citenamefont {Calci}, \citenamefont {Langhammer},\
  and\ \citenamefont {Roth}}]{Bogner:2014baa}%
  \BibitemOpen
  \bibfield  {author} {\bibinfo {author} {\bibfnamefont {S.~K.}\ \bibnamefont
  {Bogner}}, \bibinfo {author} {\bibfnamefont {H.}~\bibnamefont {Hergert}},
  \bibinfo {author} {\bibfnamefont {J.~D.}\ \bibnamefont {Holt}}, \bibinfo
  {author} {\bibfnamefont {A.}~\bibnamefont {Schwenk}}, \bibinfo {author}
  {\bibfnamefont {S.}~\bibnamefont {Binder}}, \bibinfo {author} {\bibfnamefont
  {A.}~\bibnamefont {Calci}}, \bibinfo {author} {\bibfnamefont
  {J.}~\bibnamefont {Langhammer}}, \ and\ \bibinfo {author} {\bibfnamefont
  {R.}~\bibnamefont {Roth}},\ }\href {\doibase 10.1103/PhysRevLett.113.142501}
  {\bibfield  {journal} {\bibinfo  {journal} {Phys. Rev. Lett.}\ }\textbf
  {\bibinfo {volume} {113}},\ \bibinfo {pages} {142501} (\bibinfo {year}
  {2014})},\ \Eprint {http://arxiv.org/abs/1402.1407} {arXiv:1402.1407
  [nucl-th]} \BibitemShut {NoStop}%
\bibitem [{\citenamefont {Jansen}\ \emph {et~al.}(2014)\citenamefont {Jansen},
  \citenamefont {Engel}, \citenamefont {Hagen}, \citenamefont {Navr{\`a}til},\
  and\ \citenamefont {Signoracci}}]{Jansen:2014qxa}%
  \BibitemOpen
  \bibfield  {author} {\bibinfo {author} {\bibfnamefont {G.~R.}\ \bibnamefont
  {Jansen}}, \bibinfo {author} {\bibfnamefont {J.}~\bibnamefont {Engel}},
  \bibinfo {author} {\bibfnamefont {G.}~\bibnamefont {Hagen}}, \bibinfo
  {author} {\bibfnamefont {P.}~\bibnamefont {Navr{\`a}til}}, \ and\ \bibinfo
  {author} {\bibfnamefont {A.}~\bibnamefont {Signoracci}},\ }\href {\doibase
  10.1103/PhysRevLett.113.142502} {\bibfield  {journal} {\bibinfo  {journal}
  {Phys. Rev. Lett.}\ }\textbf {\bibinfo {volume} {113}},\ \bibinfo {pages}
  {142502} (\bibinfo {year} {2014})},\ \Eprint {http://arxiv.org/abs/1402.2563}
  {arXiv:1402.2563 [nucl-th]} \BibitemShut {NoStop}%
\bibitem [{\citenamefont {Jansen}\ \emph {et~al.}(2016)\citenamefont {Jansen},
  \citenamefont {Schuster}, \citenamefont {Signoracci}, \citenamefont {Hagen},\
  and\ \citenamefont {Navr{\`a}til}}]{Jansen:2015ngw}%
  \BibitemOpen
  \bibfield  {author} {\bibinfo {author} {\bibfnamefont {G.~R.}\ \bibnamefont
  {Jansen}}, \bibinfo {author} {\bibfnamefont {M.~D.}\ \bibnamefont
  {Schuster}}, \bibinfo {author} {\bibfnamefont {A.}~\bibnamefont
  {Signoracci}}, \bibinfo {author} {\bibfnamefont {G.}~\bibnamefont {Hagen}}, \
  and\ \bibinfo {author} {\bibfnamefont {P.}~\bibnamefont {Navr{\`a}til}},\
  }\href {\doibase 10.1103/PhysRevC.94.011301} {\bibfield  {journal} {\bibinfo
  {journal} {Phys. Rev.}\ }\textbf {\bibinfo {volume} {C94}},\ \bibinfo {pages}
  {011301(R)} (\bibinfo {year} {2016})},\ \Eprint
  {http://arxiv.org/abs/1511.00757} {arXiv:1511.00757 [nucl-th]} \BibitemShut
  {NoStop}%
\bibitem [{\citenamefont {Stroberg}\ \emph {et~al.}(2016)\citenamefont
  {Stroberg}, \citenamefont {Hergert}, \citenamefont {Holt}, \citenamefont
  {Bogner},\ and\ \citenamefont {Schwenk}}]{Stroberg:2015ymf}%
  \BibitemOpen
  \bibfield  {author} {\bibinfo {author} {\bibfnamefont {S.~R.}\ \bibnamefont
  {Stroberg}}, \bibinfo {author} {\bibfnamefont {H.}~\bibnamefont {Hergert}},
  \bibinfo {author} {\bibfnamefont {J.~D.}\ \bibnamefont {Holt}}, \bibinfo
  {author} {\bibfnamefont {S.~K.}\ \bibnamefont {Bogner}}, \ and\ \bibinfo
  {author} {\bibfnamefont {A.}~\bibnamefont {Schwenk}},\ }\href {\doibase
  10.1103/PhysRevC.93.051301} {\bibfield  {journal} {\bibinfo  {journal} {Phys.
  Rev.}\ }\textbf {\bibinfo {volume} {C93}},\ \bibinfo {pages} {051301(R)}
  (\bibinfo {year} {2016})},\ \Eprint {http://arxiv.org/abs/1511.02802}
  {arXiv:1511.02802 [nucl-th]} \BibitemShut {NoStop}%
\bibitem [{\citenamefont {Stroberg}\ \emph {et~al.}(2017)\citenamefont
  {Stroberg}, \citenamefont {Calci}, \citenamefont {Hergert}, \citenamefont
  {Holt}, \citenamefont {Bogner}, \citenamefont {Roth},\ and\ \citenamefont
  {Schwenk}}]{Stroberg:2016ung}%
  \BibitemOpen
  \bibfield  {author} {\bibinfo {author} {\bibfnamefont {S.~R.}\ \bibnamefont
  {Stroberg}}, \bibinfo {author} {\bibfnamefont {A.}~\bibnamefont {Calci}},
  \bibinfo {author} {\bibfnamefont {H.}~\bibnamefont {Hergert}}, \bibinfo
  {author} {\bibfnamefont {J.~D.}\ \bibnamefont {Holt}}, \bibinfo {author}
  {\bibfnamefont {S.~K.}\ \bibnamefont {Bogner}}, \bibinfo {author}
  {\bibfnamefont {R.}~\bibnamefont {Roth}}, \ and\ \bibinfo {author}
  {\bibfnamefont {A.}~\bibnamefont {Schwenk}},\ }\href {\doibase
  10.1103/PhysRevLett.118.032502} {\bibfield  {journal} {\bibinfo  {journal}
  {Phys. Rev. Lett.}\ }\textbf {\bibinfo {volume} {118}},\ \bibinfo {pages}
  {032502} (\bibinfo {year} {2017})},\ \Eprint
  {http://arxiv.org/abs/1607.03229} {arXiv:1607.03229 [nucl-th]} \BibitemShut
  {NoStop}%
\bibitem [{\citenamefont {Dikmen}\ \emph {et~al.}(2015)\citenamefont {Dikmen},
  \citenamefont {Lisetskiy}, \citenamefont {Barrett}, \citenamefont {Maris},
  \citenamefont {Shirokov},\ and\ \citenamefont {Vary}}]{Dikmen:2015tla}%
  \BibitemOpen
  \bibfield  {author} {\bibinfo {author} {\bibfnamefont {E.}~\bibnamefont
  {Dikmen}}, \bibinfo {author} {\bibfnamefont {A.~F.}\ \bibnamefont
  {Lisetskiy}}, \bibinfo {author} {\bibfnamefont {B.~R.}\ \bibnamefont
  {Barrett}}, \bibinfo {author} {\bibfnamefont {P.}~\bibnamefont {Maris}},
  \bibinfo {author} {\bibfnamefont {A.~M.}\ \bibnamefont {Shirokov}}, \ and\
  \bibinfo {author} {\bibfnamefont {J.~P.}\ \bibnamefont {Vary}},\ }\href
  {\doibase 10.1103/PhysRevC.91.064301} {\bibfield  {journal} {\bibinfo
  {journal} {Phys. Rev.}\ }\textbf {\bibinfo {volume} {C91}},\ \bibinfo {pages}
  {064301} (\bibinfo {year} {2015})},\ \Eprint
  {http://arxiv.org/abs/1502.00700} {arXiv:1502.00700 [nucl-th]} \BibitemShut
  {NoStop}%
\bibitem [{\citenamefont {Barrett}\ \emph {et~al.}(2017)\citenamefont
  {Barrett}, \citenamefont {Dikmen}, \citenamefont {Maris}, \citenamefont
  {Shirokov}, \citenamefont {Smirnova},\ and\ \citenamefont
  {Vary}}]{Barrett:2017ovf}%
  \BibitemOpen
  \bibfield  {author} {\bibinfo {author} {\bibfnamefont {B.~R.}\ \bibnamefont
  {Barrett}}, \bibinfo {author} {\bibfnamefont {E.}~\bibnamefont {Dikmen}},
  \bibinfo {author} {\bibfnamefont {P.}~\bibnamefont {Maris}}, \bibinfo
  {author} {\bibfnamefont {A.~M.}\ \bibnamefont {Shirokov}}, \bibinfo {author}
  {\bibfnamefont {N.~A.}\ \bibnamefont {Smirnova}}, \ and\ \bibinfo {author}
  {\bibfnamefont {J.~P.}\ \bibnamefont {Vary}},\ }\bibfield  {booktitle} {\emph
  {\bibinfo {booktitle} {{Proceedings, 14th International Symposium on Nuclei
  in the Cosmos (NIC-XIV): Niigata, Japan, June 19-24, 2016}}},\ }\href
  {\doibase 10.7566/JPSCP.14.021006} {\bibfield  {journal} {\bibinfo  {journal}
  {JPS Conf. Proc.}\ }\textbf {\bibinfo {volume} {14}},\ \bibinfo {pages}
  {021006} (\bibinfo {year} {2017})}\BibitemShut {NoStop}%
\bibitem [{\citenamefont {Song}\ \emph {et~al.}(2017)\citenamefont {Song},
  \citenamefont {Yao}, \citenamefont {Ring},\ and\ \citenamefont
  {Meng}}]{Song:2017ktj}%
  \BibitemOpen
  \bibfield  {author} {\bibinfo {author} {\bibfnamefont {L.~S.}\ \bibnamefont
  {Song}}, \bibinfo {author} {\bibfnamefont {J.~M.}\ \bibnamefont {Yao}},
  \bibinfo {author} {\bibfnamefont {P.}~\bibnamefont {Ring}}, \ and\ \bibinfo
  {author} {\bibfnamefont {J.}~\bibnamefont {Meng}},\ }\href {\doibase
  10.1103/PhysRevC.95.024305} {\bibfield  {journal} {\bibinfo  {journal} {Phys.
  Rev.}\ }\textbf {\bibinfo {volume} {C95}},\ \bibinfo {pages} {024305}
  (\bibinfo {year} {2017})},\ \Eprint {http://arxiv.org/abs/1702.02448}
  {arXiv:1702.02448 [nucl-th]} \BibitemShut {NoStop}%
\bibitem [{\citenamefont {Yao}\ \emph {et~al.}(2018)\citenamefont {Yao},
  \citenamefont {Engel}, \citenamefont {Wang}, \citenamefont {Jiao},\ and\
  \citenamefont {Hergert}}]{Yao:2018qjv}%
  \BibitemOpen
  \bibfield  {author} {\bibinfo {author} {\bibfnamefont {J.~M.}\ \bibnamefont
  {Yao}}, \bibinfo {author} {\bibfnamefont {J.}~\bibnamefont {Engel}}, \bibinfo
  {author} {\bibfnamefont {L.~J.}\ \bibnamefont {Wang}}, \bibinfo {author}
  {\bibfnamefont {C.~F.}\ \bibnamefont {Jiao}}, \ and\ \bibinfo {author}
  {\bibfnamefont {H.}~\bibnamefont {Hergert}},\ }\href {\doibase
  10.1103/PhysRevC.98.054311} {\bibfield  {journal} {\bibinfo  {journal} {Phys.
  Rev.}\ }\textbf {\bibinfo {volume} {C98}},\ \bibinfo {pages} {054311}
  (\bibinfo {year} {2018})},\ \Eprint {http://arxiv.org/abs/1807.11053}
  {arXiv:1807.11053 [nucl-th]} \BibitemShut {NoStop}%
\bibitem [{\citenamefont {Entem}\ and\ \citenamefont
  {Machleidt}(2003)}]{Entem:2003ft}%
  \BibitemOpen
  \bibfield  {author} {\bibinfo {author} {\bibfnamefont {D.~R.}\ \bibnamefont
  {Entem}}\ and\ \bibinfo {author} {\bibfnamefont {R.}~\bibnamefont
  {Machleidt}},\ }\href {\doibase 10.1103/PhysRevC.68.041001} {\bibfield
  {journal} {\bibinfo  {journal} {Phys. Rev.}\ }\textbf {\bibinfo {volume}
  {C68}},\ \bibinfo {pages} {041001(R)} (\bibinfo {year} {2003})},\ \Eprint
  {http://arxiv.org/abs/nucl-th/0304018} {arXiv:nucl-th/0304018 [nucl-th]}
  \BibitemShut {NoStop}%
\bibitem [{\citenamefont {Machleidt}\ and\ \citenamefont
  {Entem}(2011)}]{Machleidt:2011zz}%
  \BibitemOpen
  \bibfield  {author} {\bibinfo {author} {\bibfnamefont {R.}~\bibnamefont
  {Machleidt}}\ and\ \bibinfo {author} {\bibfnamefont {D.~R.}\ \bibnamefont
  {Entem}},\ }\href {\doibase 10.1016/j.physrep.2011.02.001} {\bibfield
  {journal} {\bibinfo  {journal} {Phys. Rept.}\ }\textbf {\bibinfo {volume}
  {503}},\ \bibinfo {pages} {1} (\bibinfo {year} {2011})},\ \Eprint
  {http://arxiv.org/abs/1105.2919} {arXiv:1105.2919 [nucl-th]} \BibitemShut
  {NoStop}%
\bibitem [{\citenamefont {Bogner}\ \emph {et~al.}(2010)\citenamefont {Bogner},
  \citenamefont {Furnstahl},\ and\ \citenamefont {Schwenk}}]{Bogner:2009bt}%
  \BibitemOpen
  \bibfield  {author} {\bibinfo {author} {\bibfnamefont {S.~K.}\ \bibnamefont
  {Bogner}}, \bibinfo {author} {\bibfnamefont {R.~J.}\ \bibnamefont
  {Furnstahl}}, \ and\ \bibinfo {author} {\bibfnamefont {A.}~\bibnamefont
  {Schwenk}},\ }\href {\doibase 10.1016/j.ppnp.2010.03.001} {\bibfield
  {journal} {\bibinfo  {journal} {Prog. Part. Nucl. Phys.}\ }\textbf {\bibinfo
  {volume} {65}},\ \bibinfo {pages} {94} (\bibinfo {year} {2010})},\ \Eprint
  {http://arxiv.org/abs/0912.3688} {arXiv:0912.3688 [nucl-th]} \BibitemShut
  {NoStop}%
\bibitem [{\citenamefont {Cockrell}\ \emph {et~al.}(2012)\citenamefont
  {Cockrell}, \citenamefont {Vary},\ and\ \citenamefont
  {Maris}}]{Cockrell:2012vd}%
  \BibitemOpen
  \bibfield  {author} {\bibinfo {author} {\bibfnamefont {C.}~\bibnamefont
  {Cockrell}}, \bibinfo {author} {\bibfnamefont {J.~P.}\ \bibnamefont {Vary}},
  \ and\ \bibinfo {author} {\bibfnamefont {P.}~\bibnamefont {Maris}},\ }\href
  {\doibase 10.1103/PhysRevC.86.034325} {\bibfield  {journal} {\bibinfo
  {journal} {Phys. Rev.}\ }\textbf {\bibinfo {volume} {C86}},\ \bibinfo {pages}
  {034325} (\bibinfo {year} {2012})},\ \Eprint {http://arxiv.org/abs/1201.0724}
  {arXiv:1201.0724 [nucl-th]} \BibitemShut {NoStop}%
\bibitem [{\citenamefont {Shin}\ \emph {et~al.}(2017)\citenamefont {Shin},
  \citenamefont {Kim}, \citenamefont {Maris}, \citenamefont {Vary},
  \citenamefont {Forss{\`e}n}, \citenamefont {Rotureau},\ and\ \citenamefont
  {Michel}}]{Shin:2016poa}%
  \BibitemOpen
  \bibfield  {author} {\bibinfo {author} {\bibfnamefont {I.~J.}\ \bibnamefont
  {Shin}}, \bibinfo {author} {\bibfnamefont {Y.}~\bibnamefont {Kim}}, \bibinfo
  {author} {\bibfnamefont {P.}~\bibnamefont {Maris}}, \bibinfo {author}
  {\bibfnamefont {J.~P.}\ \bibnamefont {Vary}}, \bibinfo {author}
  {\bibfnamefont {C.}~\bibnamefont {Forss{\`e}n}}, \bibinfo {author}
  {\bibfnamefont {J.}~\bibnamefont {Rotureau}}, \ and\ \bibinfo {author}
  {\bibfnamefont {N.}~\bibnamefont {Michel}},\ }\href {\doibase
  10.1088/1361-6471/aa6cb7} {\bibfield  {journal} {\bibinfo  {journal} {J.
  Phys.}\ }\textbf {\bibinfo {volume} {G44}},\ \bibinfo {pages} {075103}
  (\bibinfo {year} {2017})},\ \Eprint {http://arxiv.org/abs/1605.02819}
  {arXiv:1605.02819 [nucl-th]} \BibitemShut {NoStop}%
\bibitem [{\citenamefont {Pastore}\ \emph {et~al.}(2018)\citenamefont
  {Pastore}, \citenamefont {Carlson}, \citenamefont {Cirigliano}, \citenamefont
  {Dekens}, \citenamefont {Mereghetti},\ and\ \citenamefont
  {Wiringa}}]{Pastore:2017ofx}%
  \BibitemOpen
  \bibfield  {author} {\bibinfo {author} {\bibfnamefont {S.}~\bibnamefont
  {Pastore}}, \bibinfo {author} {\bibfnamefont {J.}~\bibnamefont {Carlson}},
  \bibinfo {author} {\bibfnamefont {V.}~\bibnamefont {Cirigliano}}, \bibinfo
  {author} {\bibfnamefont {W.}~\bibnamefont {Dekens}}, \bibinfo {author}
  {\bibfnamefont {E.}~\bibnamefont {Mereghetti}}, \ and\ \bibinfo {author}
  {\bibfnamefont {R.~B.}\ \bibnamefont {Wiringa}},\ }\href {\doibase
  10.1103/PhysRevC.97.014606} {\bibfield  {journal} {\bibinfo  {journal} {Phys.
  Rev.}\ }\textbf {\bibinfo {volume} {C97}},\ \bibinfo {pages} {014606}
  (\bibinfo {year} {2018})},\ \Eprint {http://arxiv.org/abs/1710.05026}
  {arXiv:1710.05026 [nucl-th]} \BibitemShut {NoStop}%
\bibitem [{\citenamefont {Avignone}\ \emph {et~al.}(2008)\citenamefont
  {Avignone}, \citenamefont {Elliott},\ and\ \citenamefont
  {Engel}}]{Avignone:2007fu}%
  \BibitemOpen
  \bibfield  {author} {\bibinfo {author} {\bibfnamefont {F.~T.}\ \bibnamefont
  {Avignone}, \bibfnamefont {III}}, \bibinfo {author} {\bibfnamefont {S.~R.}\
  \bibnamefont {Elliott}}, \ and\ \bibinfo {author} {\bibfnamefont
  {J.}~\bibnamefont {Engel}},\ }\href {\doibase 10.1103/RevModPhys.80.481}
  {\bibfield  {journal} {\bibinfo  {journal} {Rev. Mod. Phys.}\ }\textbf
  {\bibinfo {volume} {80}},\ \bibinfo {pages} {481} (\bibinfo {year} {2008})},\
  \Eprint {http://arxiv.org/abs/0708.1033} {arXiv:0708.1033 [nucl-ex]}
  \BibitemShut {NoStop}%
\bibitem [{\citenamefont {Simkovic}\ \emph {et~al.}(1999)\citenamefont
  {Simkovic}, \citenamefont {Pantis}, \citenamefont {Vergados},\ and\
  \citenamefont {Faessler}}]{Simkovic:1999re}%
  \BibitemOpen
  \bibfield  {author} {\bibinfo {author} {\bibfnamefont {F.}~\bibnamefont
  {Simkovic}}, \bibinfo {author} {\bibfnamefont {G.}~\bibnamefont {Pantis}},
  \bibinfo {author} {\bibfnamefont {J.~D.}\ \bibnamefont {Vergados}}, \ and\
  \bibinfo {author} {\bibfnamefont {A.}~\bibnamefont {Faessler}},\ }\href
  {\doibase 10.1103/PhysRevC.60.055502} {\bibfield  {journal} {\bibinfo
  {journal} {Phys. Rev.}\ }\textbf {\bibinfo {volume} {C60}},\ \bibinfo {pages}
  {055502} (\bibinfo {year} {1999})},\ \Eprint
  {http://arxiv.org/abs/hep-ph/9905509} {arXiv:hep-ph/9905509 [hep-ph]}
  \BibitemShut {NoStop}%
\bibitem [{\citenamefont {Kotila}\ and\ \citenamefont
  {Iachello}(2012)}]{Kotila:2012zza}%
  \BibitemOpen
  \bibfield  {author} {\bibinfo {author} {\bibfnamefont {J.}~\bibnamefont
  {Kotila}}\ and\ \bibinfo {author} {\bibfnamefont {F.}~\bibnamefont
  {Iachello}},\ }\href {\doibase 10.1103/PhysRevC.85.034316} {\bibfield
  {journal} {\bibinfo  {journal} {Phys. Rev.}\ }\textbf {\bibinfo {volume}
  {C85}},\ \bibinfo {pages} {034316} (\bibinfo {year} {2012})},\ \Eprint
  {http://arxiv.org/abs/1209.5722} {arXiv:1209.5722 [nucl-th]} \BibitemShut
  {NoStop}%
\bibitem [{\citenamefont {Stoica}\ and\ \citenamefont
  {Mirea}(2013)}]{Stoica:2013lka}%
  \BibitemOpen
  \bibfield  {author} {\bibinfo {author} {\bibfnamefont {S.}~\bibnamefont
  {Stoica}}\ and\ \bibinfo {author} {\bibfnamefont {M.}~\bibnamefont {Mirea}},\
  }\href {\doibase 10.1103/PhysRevC.88.037303} {\bibfield  {journal} {\bibinfo
  {journal} {Phys. Rev.}\ }\textbf {\bibinfo {volume} {C88}},\ \bibinfo {pages}
  {037303} (\bibinfo {year} {2013})},\ \Eprint {http://arxiv.org/abs/1307.0290}
  {arXiv:1307.0290 [nucl-th]} \BibitemShut {NoStop}%
\bibitem [{\citenamefont {Horoi}\ and\ \citenamefont
  {Stoica}(2010)}]{Horoi:2009gz}%
  \BibitemOpen
  \bibfield  {author} {\bibinfo {author} {\bibfnamefont {M.}~\bibnamefont
  {Horoi}}\ and\ \bibinfo {author} {\bibfnamefont {S.}~\bibnamefont {Stoica}},\
  }\href {\doibase 10.1103/PhysRevC.81.024321} {\bibfield  {journal} {\bibinfo
  {journal} {Phys. Rev.}\ }\textbf {\bibinfo {volume} {C81}},\ \bibinfo {pages}
  {024321} (\bibinfo {year} {2010})},\ \Eprint {http://arxiv.org/abs/0911.3807}
  {arXiv:0911.3807 [nucl-th]} \BibitemShut {NoStop}%
\bibitem [{\citenamefont {Rodin}\ \emph {et~al.}(2006)\citenamefont {Rodin},
  \citenamefont {Faessler}, \citenamefont {Simkovic},\ and\ \citenamefont
  {Vogel}}]{Rodin:2006yk}%
  \BibitemOpen
  \bibfield  {author} {\bibinfo {author} {\bibfnamefont {V.~A.}\ \bibnamefont
  {Rodin}}, \bibinfo {author} {\bibfnamefont {A.}~\bibnamefont {Faessler}},
  \bibinfo {author} {\bibfnamefont {F.}~\bibnamefont {Simkovic}}, \ and\
  \bibinfo {author} {\bibfnamefont {P.}~\bibnamefont {Vogel}},\ }\href
  {\doibase 10.1016/j.nuclphysa.2007.06.014, 10.1016/j.nuclphysa.2005.12.004}
  {\bibfield  {journal} {\bibinfo  {journal} {Nucl. Phys.}\ }\textbf {\bibinfo
  {volume} {A766}},\ \bibinfo {pages} {107} (\bibinfo {year} {2006})},\
  \bibinfo {note} {[Erratum: Nucl. Phys.A793,213(2007)]},\ \Eprint
  {http://arxiv.org/abs/0706.4304} {arXiv:0706.4304 [nucl-th]} \BibitemShut
  {NoStop}%
\bibitem [{\citenamefont {Suhonen}(2007)}]{Suhonen:2007zza}%
  \BibitemOpen
  \bibfield  {author} {\bibinfo {author} {\bibfnamefont {J.}~\bibnamefont
  {Suhonen}},\ }\href {\doibase 10.1007/978-3-540-48861-3} {\emph {\bibinfo
  {title} {{From Nucleons to Nucleus}}}},\ Theoretical and Mathematical
  Physics\ (\bibinfo  {publisher} {Springer},\ \bibinfo {address} {Berlin,
  Germany},\ \bibinfo {year} {2007})\BibitemShut {NoStop}%
\bibitem [{\citenamefont {Simkovic}\ \emph {et~al.}(2008)\citenamefont
  {Simkovic}, \citenamefont {Faessler}, \citenamefont {Rodin}, \citenamefont
  {Vogel},\ and\ \citenamefont {Engel}}]{Simkovic:2007vu}%
  \BibitemOpen
  \bibfield  {author} {\bibinfo {author} {\bibfnamefont {F.}~\bibnamefont
  {Simkovic}}, \bibinfo {author} {\bibfnamefont {A.}~\bibnamefont {Faessler}},
  \bibinfo {author} {\bibfnamefont {V.}~\bibnamefont {Rodin}}, \bibinfo
  {author} {\bibfnamefont {P.}~\bibnamefont {Vogel}}, \ and\ \bibinfo {author}
  {\bibfnamefont {J.}~\bibnamefont {Engel}},\ }\href {\doibase
  10.1103/PhysRevC.77.045503} {\bibfield  {journal} {\bibinfo  {journal} {Phys.
  Rev.}\ }\textbf {\bibinfo {volume} {C77}},\ \bibinfo {pages} {045503}
  (\bibinfo {year} {2008})},\ \Eprint {http://arxiv.org/abs/0710.2055}
  {arXiv:0710.2055 [nucl-th]} \BibitemShut {NoStop}%
\bibitem [{\citenamefont {Simkovic}\ \emph {et~al.}(2009)\citenamefont
  {Simkovic}, \citenamefont {Faessler}, \citenamefont {Muther}, \citenamefont
  {Rodin},\ and\ \citenamefont {Stauf}}]{Simkovic:2009pp}%
  \BibitemOpen
  \bibfield  {author} {\bibinfo {author} {\bibfnamefont {F.}~\bibnamefont
  {Simkovic}}, \bibinfo {author} {\bibfnamefont {A.}~\bibnamefont {Faessler}},
  \bibinfo {author} {\bibfnamefont {H.}~\bibnamefont {Muther}}, \bibinfo
  {author} {\bibfnamefont {V.}~\bibnamefont {Rodin}}, \ and\ \bibinfo {author}
  {\bibfnamefont {M.}~\bibnamefont {Stauf}},\ }\href {\doibase
  10.1103/PhysRevC.79.055501} {\bibfield  {journal} {\bibinfo  {journal} {Phys.
  Rev.}\ }\textbf {\bibinfo {volume} {C79}},\ \bibinfo {pages} {055501}
  (\bibinfo {year} {2009})},\ \Eprint {http://arxiv.org/abs/0902.0331}
  {arXiv:0902.0331 [nucl-th]} \BibitemShut {NoStop}%
\bibitem [{\citenamefont {Miller}\ and\ \citenamefont
  {Spencer}(1976)}]{Miller:1975hu}%
  \BibitemOpen
  \bibfield  {author} {\bibinfo {author} {\bibfnamefont {G.~A.}\ \bibnamefont
  {Miller}}\ and\ \bibinfo {author} {\bibfnamefont {J.~E.}\ \bibnamefont
  {Spencer}},\ }\href {\doibase 10.1016/0003-4916(76)90073-7} {\bibfield
  {journal} {\bibinfo  {journal} {Annals Phys.}\ }\textbf {\bibinfo {volume}
  {100}},\ \bibinfo {pages} {562} (\bibinfo {year} {1976})}\BibitemShut
  {NoStop}%
\bibitem [{\citenamefont {Roth}\ \emph {et~al.}(2005)\citenamefont {Roth},
  \citenamefont {Hergert}, \citenamefont {Papakonstantinou}, \citenamefont
  {Neff},\ and\ \citenamefont {Feldmeier}}]{Roth:2005pd}%
  \BibitemOpen
  \bibfield  {author} {\bibinfo {author} {\bibfnamefont {R.}~\bibnamefont
  {Roth}}, \bibinfo {author} {\bibfnamefont {H.}~\bibnamefont {Hergert}},
  \bibinfo {author} {\bibfnamefont {P.}~\bibnamefont {Papakonstantinou}},
  \bibinfo {author} {\bibfnamefont {T.}~\bibnamefont {Neff}}, \ and\ \bibinfo
  {author} {\bibfnamefont {H.}~\bibnamefont {Feldmeier}},\ }\href {\doibase
  10.1103/PhysRevC.72.034002} {\bibfield  {journal} {\bibinfo  {journal} {Phys.
  Rev.}\ }\textbf {\bibinfo {volume} {C72}},\ \bibinfo {pages} {034002}
  (\bibinfo {year} {2005})},\ \Eprint {http://arxiv.org/abs/nucl-th/0505080}
  {arXiv:nucl-th/0505080 [nucl-th]} \BibitemShut {NoStop}%
\bibitem [{\citenamefont {Brueckner}(1955)}]{Brueckner:1955zza}%
  \BibitemOpen
  \bibfield  {author} {\bibinfo {author} {\bibfnamefont {K.~A.}\ \bibnamefont
  {Brueckner}},\ }\href {\doibase 10.1103/PhysRev.100.36} {\bibfield  {journal}
  {\bibinfo  {journal} {Phys. Rev.}\ }\textbf {\bibinfo {volume} {100}},\
  \bibinfo {pages} {36} (\bibinfo {year} {1955})}\BibitemShut {NoStop}%
\bibitem [{\citenamefont {Benhar}\ \emph {et~al.}(2014)\citenamefont {Benhar},
  \citenamefont {Biondi},\ and\ \citenamefont {Speranza}}]{Benhar:2014cka}%
  \BibitemOpen
  \bibfield  {author} {\bibinfo {author} {\bibfnamefont {O.}~\bibnamefont
  {Benhar}}, \bibinfo {author} {\bibfnamefont {R.}~\bibnamefont {Biondi}}, \
  and\ \bibinfo {author} {\bibfnamefont {E.}~\bibnamefont {Speranza}},\ }\href
  {\doibase 10.1103/PhysRevC.90.065504} {\bibfield  {journal} {\bibinfo
  {journal} {Phys. Rev.}\ }\textbf {\bibinfo {volume} {C90}},\ \bibinfo {pages}
  {065504} (\bibinfo {year} {2014})},\ \Eprint {http://arxiv.org/abs/1401.2030}
  {arXiv:1401.2030 [nucl-th]} \BibitemShut {NoStop}%
\bibitem [{\citenamefont {Muther}\ and\ \citenamefont
  {Polls}(2000)}]{Muther:2000qx}%
  \BibitemOpen
  \bibfield  {author} {\bibinfo {author} {\bibfnamefont {H.}~\bibnamefont
  {Muther}}\ and\ \bibinfo {author} {\bibfnamefont {A.}~\bibnamefont {Polls}},\
  }\href {\doibase 10.1016/S0146-6410(00)00105-8} {\bibfield  {journal}
  {\bibinfo  {journal} {Prog. Part. Nucl. Phys.}\ }\textbf {\bibinfo {volume}
  {45}},\ \bibinfo {pages} {243} (\bibinfo {year} {2000})},\ \Eprint
  {http://arxiv.org/abs/nucl-th/0001007} {arXiv:nucl-th/0001007 [nucl-th]}
  \BibitemShut {NoStop}%
\bibitem [{\citenamefont {Tobocman}(1981)}]{Tobocman:1981yao}%
  \BibitemOpen
  \bibfield  {author} {\bibinfo {author} {\bibfnamefont {W.}~\bibnamefont
  {Tobocman}},\ }\href {\doibase 10.1016/0375-9474(81)90223-2} {\bibfield
  {journal} {\bibinfo  {journal} {Nucl. Phys.}\ }\textbf {\bibinfo {volume}
  {A357}},\ \bibinfo {pages} {293} (\bibinfo {year} {1981})}\BibitemShut
  {NoStop}%
\bibitem [{\citenamefont {Barrett}\ \emph {et~al.}(2013)\citenamefont
  {Barrett}, \citenamefont {Navr{\`a}til},\ and\ \citenamefont
  {Vary}}]{Barrett:2013nh}%
  \BibitemOpen
  \bibfield  {author} {\bibinfo {author} {\bibfnamefont {B.~R.}\ \bibnamefont
  {Barrett}}, \bibinfo {author} {\bibfnamefont {P.}~\bibnamefont
  {Navr{\`a}til}}, \ and\ \bibinfo {author} {\bibfnamefont {J.~P.}\
  \bibnamefont {Vary}},\ }\href {\doibase 10.1016/j.ppnp.2012.10.003}
  {\bibfield  {journal} {\bibinfo  {journal} {Prog. Part. Nucl. Phys.}\
  }\textbf {\bibinfo {volume} {69}},\ \bibinfo {pages} {131} (\bibinfo {year}
  {2013})}\BibitemShut {NoStop}%
\bibitem [{\citenamefont {Maris}\ \emph {et~al.}(2010)\citenamefont {Maris},
  \citenamefont {Sosonkina}, \citenamefont {Vary}, \citenamefont {Ng},\ and\
  \citenamefont {Yang}}]{MARIS201097}%
  \BibitemOpen
  \bibfield  {author} {\bibinfo {author} {\bibfnamefont {P.}~\bibnamefont
  {Maris}}, \bibinfo {author} {\bibfnamefont {M.}~\bibnamefont {Sosonkina}},
  \bibinfo {author} {\bibfnamefont {J.~P.}\ \bibnamefont {Vary}}, \bibinfo
  {author} {\bibfnamefont {E.}~\bibnamefont {Ng}}, \ and\ \bibinfo {author}
  {\bibfnamefont {C.}~\bibnamefont {Yang}},\ }\href {\doibase
  https://doi.org/10.1016/j.procs.2010.04.012} {\bibfield  {journal} {\bibinfo
  {journal} {Procedia Computer Science}\ }\textbf {\bibinfo {volume} {1}},\
  \bibinfo {pages} {97 } (\bibinfo {year} {2010})},\ \bibinfo {note} {iCCS
  2010}\BibitemShut {NoStop}%
\bibitem [{\citenamefont {Aktulga}\ \emph {et~al.}(2014)\citenamefont
  {Aktulga}, \citenamefont {Yang}, \citenamefont {Ng}, \citenamefont {Maris},\
  and\ \citenamefont {Vary}}]{Aktulga:2014mfd}%
  \BibitemOpen
  \bibfield  {author} {\bibinfo {author} {\bibfnamefont {H.~M.}\ \bibnamefont
  {Aktulga}}, \bibinfo {author} {\bibfnamefont {C.}~\bibnamefont {Yang}},
  \bibinfo {author} {\bibfnamefont {E.~G.}\ \bibnamefont {Ng}}, \bibinfo
  {author} {\bibfnamefont {P.}~\bibnamefont {Maris}}, \ and\ \bibinfo {author}
  {\bibfnamefont {J.~P.}\ \bibnamefont {Vary}},\ }\bibfield  {booktitle} {\emph
  {\bibinfo {booktitle} {Concurrency and Computation: Practice and
  Experience}},\ }\href@noop {} {\ \textbf {\bibinfo {volume} {26}},\ \bibinfo
  {pages} {2631} (\bibinfo {year} {2014})}\BibitemShut {NoStop}%
\bibitem [{\citenamefont {Shao}\ \emph {et~al.}(2017)\citenamefont {Shao},
  \citenamefont {Aktulga}, \citenamefont {Yang}, \citenamefont {Ng},
  \citenamefont {Maris},\ and\ \citenamefont
  {Vary}}]{DBLP:journals/corr/ShaoAYNMV16}%
  \BibitemOpen
  \bibfield  {author} {\bibinfo {author} {\bibfnamefont {M.}~\bibnamefont
  {Shao}}, \bibinfo {author} {\bibfnamefont {H.~M.}\ \bibnamefont {Aktulga}},
  \bibinfo {author} {\bibfnamefont {C.}~\bibnamefont {Yang}}, \bibinfo {author}
  {\bibfnamefont {E.~G.}\ \bibnamefont {Ng}}, \bibinfo {author} {\bibfnamefont
  {P.}~\bibnamefont {Maris}}, \ and\ \bibinfo {author} {\bibfnamefont {J.~P.}\
  \bibnamefont {Vary}},\ }\href {\doibase 10.1016/j.cpc.2017.09.004} {\bibfield
   {journal} {\bibinfo  {journal} {Computer Physics Communications}\ }\textbf
  {\bibinfo {volume} {222}} (\bibinfo {year} {2017}),\
  10.1016/j.cpc.2017.09.004},\ \Eprint {http://arxiv.org/abs/1609.01689}
  {arXiv:1609.01689} \BibitemShut {NoStop}%
\bibitem [{\citenamefont {Maris}\ \emph {et~al.}(2009)\citenamefont {Maris},
  \citenamefont {Vary},\ and\ \citenamefont {Shirokov}}]{Maris:2008ax}%
  \BibitemOpen
  \bibfield  {author} {\bibinfo {author} {\bibfnamefont {P.}~\bibnamefont
  {Maris}}, \bibinfo {author} {\bibfnamefont {J.~P.}\ \bibnamefont {Vary}}, \
  and\ \bibinfo {author} {\bibfnamefont {A.~M.}\ \bibnamefont {Shirokov}},\
  }\href {\doibase 10.1103/PhysRevC.79.014308} {\bibfield  {journal} {\bibinfo
  {journal} {Phys. Rev.}\ }\textbf {\bibinfo {volume} {C79}},\ \bibinfo {pages}
  {014308} (\bibinfo {year} {2009})},\ \Eprint {http://arxiv.org/abs/0808.3420}
  {arXiv:0808.3420 [nucl-th]} \BibitemShut {NoStop}%
\bibitem [{\citenamefont {Hergert}\ \emph {et~al.}(2016)\citenamefont
  {Hergert}, \citenamefont {Bogner}, \citenamefont {Morris}, \citenamefont
  {Schwenk},\ and\ \citenamefont {Tsukiyama}}]{Hergert:2015awm}%
  \BibitemOpen
  \bibfield  {author} {\bibinfo {author} {\bibfnamefont {H.}~\bibnamefont
  {Hergert}}, \bibinfo {author} {\bibfnamefont {S.~K.}\ \bibnamefont {Bogner}},
  \bibinfo {author} {\bibfnamefont {T.~D.}\ \bibnamefont {Morris}}, \bibinfo
  {author} {\bibfnamefont {A.}~\bibnamefont {Schwenk}}, \ and\ \bibinfo
  {author} {\bibfnamefont {K.}~\bibnamefont {Tsukiyama}},\ }\href {\doibase
  10.1016/j.physrep.2015.12.007} {\bibfield  {journal} {\bibinfo  {journal}
  {Phys. Rept.}\ }\textbf {\bibinfo {volume} {621}},\ \bibinfo {pages} {165}
  (\bibinfo {year} {2016})},\ \Eprint {http://arxiv.org/abs/1512.06956}
  {arXiv:1512.06956 [nucl-th]} \BibitemShut {NoStop}%
\bibitem [{\citenamefont {Hergert}(2017)}]{Hergert:2016etg}%
  \BibitemOpen
  \bibfield  {author} {\bibinfo {author} {\bibfnamefont {H.}~\bibnamefont
  {Hergert}},\ }\href {\doibase 10.1088/1402-4896/92/2/023002} {\bibfield
  {journal} {\bibinfo  {journal} {Phys. Scripta}\ }\textbf {\bibinfo {volume}
  {92}},\ \bibinfo {pages} {023002} (\bibinfo {year} {2017})},\ \Eprint
  {http://arxiv.org/abs/1607.06882} {arXiv:1607.06882 [nucl-th]} \BibitemShut
  {NoStop}%
\bibitem [{\citenamefont {Hergert}\ \emph {et~al.}(2018)\citenamefont
  {Hergert}, \citenamefont {Yao}, \citenamefont {Morris}, \citenamefont
  {Parzuchowski}, \citenamefont {Bogner},\ and\ \citenamefont
  {Engel}}]{Hergert:2018wmx}%
  \BibitemOpen
  \bibfield  {author} {\bibinfo {author} {\bibfnamefont {H.}~\bibnamefont
  {Hergert}}, \bibinfo {author} {\bibfnamefont {J.~M.}\ \bibnamefont {Yao}},
  \bibinfo {author} {\bibfnamefont {T.~D.}\ \bibnamefont {Morris}}, \bibinfo
  {author} {\bibfnamefont {N.~M.}\ \bibnamefont {Parzuchowski}}, \bibinfo
  {author} {\bibfnamefont {S.~K.}\ \bibnamefont {Bogner}}, \ and\ \bibinfo
  {author} {\bibfnamefont {J.}~\bibnamefont {Engel}},\ }\href@noop {}
  {\bibfield  {journal} {\bibinfo  {journal} {Journal of Physics: Conference
  Series}\ }\textbf {\bibinfo {volume} {1041}},\ \bibinfo {pages} {012007}
  (\bibinfo {year} {2018})}\BibitemShut {NoStop}%
\bibitem [{\citenamefont {Morris}\ \emph {et~al.}(2015)\citenamefont {Morris},
  \citenamefont {Parzuchowski},\ and\ \citenamefont {Bogner}}]{Morris:2015yna}%
  \BibitemOpen
  \bibfield  {author} {\bibinfo {author} {\bibfnamefont {T.~D.}\ \bibnamefont
  {Morris}}, \bibinfo {author} {\bibfnamefont {N.~M.}\ \bibnamefont
  {Parzuchowski}}, \ and\ \bibinfo {author} {\bibfnamefont {S.~K.}\
  \bibnamefont {Bogner}},\ }\href {\doibase 10.1103/PhysRevC.92.034331}
  {\bibfield  {journal} {\bibinfo  {journal} {Phys. Rev.}\ }\textbf {\bibinfo
  {volume} {C92}},\ \bibinfo {pages} {034331} (\bibinfo {year} {2015})},\
  \Eprint {http://arxiv.org/abs/1507.06725} {arXiv:1507.06725 [nucl-th]}
  \BibitemShut {NoStop}%
\bibitem [{\citenamefont {Furnstahl}\ \emph {et~al.}(2012)\citenamefont
  {Furnstahl}, \citenamefont {Hagen},\ and\ \citenamefont
  {Papenbrock}}]{Furnstahl:2012qg}%
  \BibitemOpen
  \bibfield  {author} {\bibinfo {author} {\bibfnamefont {R.~J.}\ \bibnamefont
  {Furnstahl}}, \bibinfo {author} {\bibfnamefont {G.}~\bibnamefont {Hagen}}, \
  and\ \bibinfo {author} {\bibfnamefont {T.}~\bibnamefont {Papenbrock}},\
  }\href {\doibase 10.1103/PhysRevC.86.031301} {\bibfield  {journal} {\bibinfo
  {journal} {Phys. Rev.}\ }\textbf {\bibinfo {volume} {C86}},\ \bibinfo {pages}
  {031301(R)} (\bibinfo {year} {2012})},\ \Eprint
  {http://arxiv.org/abs/1207.6100} {arXiv:1207.6100 [nucl-th]} \BibitemShut
  {NoStop}%
\bibitem [{\citenamefont {Odell}\ \emph {et~al.}(2016)\citenamefont {Odell},
  \citenamefont {Papenbrock},\ and\ \citenamefont {Platter}}]{Odell:2015xlw}%
  \BibitemOpen
  \bibfield  {author} {\bibinfo {author} {\bibfnamefont {D.}~\bibnamefont
  {Odell}}, \bibinfo {author} {\bibfnamefont {T.}~\bibnamefont {Papenbrock}}, \
  and\ \bibinfo {author} {\bibfnamefont {L.}~\bibnamefont {Platter}},\ }\href
  {\doibase 10.1103/PhysRevC.93.044331} {\bibfield  {journal} {\bibinfo
  {journal} {Phys. Rev.}\ }\textbf {\bibinfo {volume} {C93}},\ \bibinfo {pages}
  {044331} (\bibinfo {year} {2016})},\ \Eprint
  {http://arxiv.org/abs/1512.04851} {arXiv:1512.04851 [nucl-th]} \BibitemShut
  {NoStop}%
\bibitem [{\citenamefont {Riisager}\ \emph {et~al.}(1990)\citenamefont
  {Riisager} \emph {et~al.}}]{Riisager:1989hv}%
  \BibitemOpen
  \bibfield  {author} {\bibinfo {author} {\bibfnamefont {K.}~\bibnamefont
  {Riisager}} \emph {et~al.} (\bibinfo {collaboration} {ISOLDE}),\ }\href
  {\doibase 10.1016/0370-2693(90)90091-J} {\bibfield  {journal} {\bibinfo
  {journal} {Phys. Lett.}\ }\textbf {\bibinfo {volume} {B235}},\ \bibinfo
  {pages} {30} (\bibinfo {year} {1990})}\BibitemShut {NoStop}%
\bibitem [{\citenamefont {Bogner}\ \emph {et~al.}(2008)\citenamefont {Bogner},
  \citenamefont {Furnstahl}, \citenamefont {Maris}, \citenamefont {Perry},
  \citenamefont {Schwenk},\ and\ \citenamefont {Vary}}]{Bogner:2007rx}%
  \BibitemOpen
  \bibfield  {author} {\bibinfo {author} {\bibfnamefont {S.~K.}\ \bibnamefont
  {Bogner}}, \bibinfo {author} {\bibfnamefont {R.~J.}\ \bibnamefont
  {Furnstahl}}, \bibinfo {author} {\bibfnamefont {P.}~\bibnamefont {Maris}},
  \bibinfo {author} {\bibfnamefont {R.~J.}\ \bibnamefont {Perry}}, \bibinfo
  {author} {\bibfnamefont {A.}~\bibnamefont {Schwenk}}, \ and\ \bibinfo
  {author} {\bibfnamefont {J.~P.}\ \bibnamefont {Vary}},\ }\href {\doibase
  10.1016/j.nuclphysa.2007.12.008} {\bibfield  {journal} {\bibinfo  {journal}
  {Nucl. Phys.}\ }\textbf {\bibinfo {volume} {A801}},\ \bibinfo {pages} {21}
  (\bibinfo {year} {2008})},\ \Eprint {http://arxiv.org/abs/0708.3754}
  {arXiv:0708.3754 [nucl-th]} \BibitemShut {NoStop}%
\bibitem [{\citenamefont {Wang}\ \emph {et~al.}(2017)\citenamefont {Wang},
  \citenamefont {Audi}, \citenamefont {Kondev}, \citenamefont {Huang},
  \citenamefont {Naimi},\ and\ \citenamefont {Xu}}]{Wang:2017}%
  \BibitemOpen
  \bibfield  {author} {\bibinfo {author} {\bibfnamefont {M.}~\bibnamefont
  {Wang}}, \bibinfo {author} {\bibfnamefont {G.}~\bibnamefont {Audi}}, \bibinfo
  {author} {\bibfnamefont {F.~G.}\ \bibnamefont {Kondev}}, \bibinfo {author}
  {\bibfnamefont {W.}~\bibnamefont {Huang}}, \bibinfo {author} {\bibfnamefont
  {S.}~\bibnamefont {Naimi}}, \ and\ \bibinfo {author} {\bibfnamefont
  {X.}~\bibnamefont {Xu}},\ }\href {\doibase 10.1088/1674-1137/41/3/030003}
  {\bibfield  {journal} {\bibinfo  {journal} {Chinese Physics C}\ }\textbf
  {\bibinfo {volume} {41}},\ \bibinfo {pages} {030003} (\bibinfo {year}
  {2017})}\BibitemShut {NoStop}%
\bibitem [{\citenamefont {Furnstahl}\ \emph {et~al.}(2014)\citenamefont
  {Furnstahl}, \citenamefont {More},\ and\ \citenamefont
  {Papenbrock}}]{Furnstahl:2013vda}%
  \BibitemOpen
  \bibfield  {author} {\bibinfo {author} {\bibfnamefont {R.~J.}\ \bibnamefont
  {Furnstahl}}, \bibinfo {author} {\bibfnamefont {S.~N.}\ \bibnamefont {More}},
  \ and\ \bibinfo {author} {\bibfnamefont {T.}~\bibnamefont {Papenbrock}},\
  }\href {\doibase 10.1103/PhysRevC.89.044301} {\bibfield  {journal} {\bibinfo
  {journal} {Phys. Rev.}\ }\textbf {\bibinfo {volume} {C89}},\ \bibinfo {pages}
  {044301} (\bibinfo {year} {2014})},\ \Eprint {http://arxiv.org/abs/1312.6876}
  {arXiv:1312.6876 [nucl-th]} \BibitemShut {NoStop}%
\bibitem [{\citenamefont {Furnstahl}\ \emph {et~al.}(2015)\citenamefont
  {Furnstahl}, \citenamefont {Hagen}, \citenamefont {Papenbrock},\ and\
  \citenamefont {Wendt}}]{Furnstahl:2014hca}%
  \BibitemOpen
  \bibfield  {author} {\bibinfo {author} {\bibfnamefont {R.~J.}\ \bibnamefont
  {Furnstahl}}, \bibinfo {author} {\bibfnamefont {G.}~\bibnamefont {Hagen}},
  \bibinfo {author} {\bibfnamefont {T.}~\bibnamefont {Papenbrock}}, \ and\
  \bibinfo {author} {\bibfnamefont {K.~A.}\ \bibnamefont {Wendt}},\ }\href
  {\doibase 10.1088/0954-3899/42/3/034032} {\bibfield  {journal} {\bibinfo
  {journal} {J. Phys.}\ }\textbf {\bibinfo {volume} {G42}},\ \bibinfo {pages}
  {034032} (\bibinfo {year} {2015})},\ \Eprint {http://arxiv.org/abs/1408.0252}
  {arXiv:1408.0252 [nucl-th]} \BibitemShut {NoStop}%
\bibitem [{\citenamefont {Coon}\ \emph {et~al.}(2012)\citenamefont {Coon},
  \citenamefont {Avetian}, \citenamefont {Kruse}, \citenamefont {van Kolck},
  \citenamefont {Maris},\ and\ \citenamefont {Vary}}]{Coon:2012ab}%
  \BibitemOpen
  \bibfield  {author} {\bibinfo {author} {\bibfnamefont {S.~A.}\ \bibnamefont
  {Coon}}, \bibinfo {author} {\bibfnamefont {M.~I.}\ \bibnamefont {Avetian}},
  \bibinfo {author} {\bibfnamefont {M.~K.~G.}\ \bibnamefont {Kruse}}, \bibinfo
  {author} {\bibfnamefont {U.}~\bibnamefont {van Kolck}}, \bibinfo {author}
  {\bibfnamefont {P.}~\bibnamefont {Maris}}, \ and\ \bibinfo {author}
  {\bibfnamefont {J.~P.}\ \bibnamefont {Vary}},\ }\href {\doibase
  10.1103/PhysRevC.86.054002} {\bibfield  {journal} {\bibinfo  {journal} {Phys.
  Rev.}\ }\textbf {\bibinfo {volume} {C86}},\ \bibinfo {pages} {054002}
  (\bibinfo {year} {2012})},\ \Eprint {http://arxiv.org/abs/1205.3230}
  {arXiv:1205.3230 [nucl-th]} \BibitemShut {NoStop}%
\bibitem [{\citenamefont {Riisager}(2013)}]{Riisager:2012it}%
  \BibitemOpen
  \bibfield  {author} {\bibinfo {author} {\bibfnamefont {K.}~\bibnamefont
  {Riisager}},\ }\href {\doibase 10.1088/0031-8949/2013/T152/014001} {\bibfield
   {journal} {\bibinfo  {journal} {Phys. Scripta}\ }\textbf {\bibinfo {volume}
  {T152}},\ \bibinfo {pages} {014001} (\bibinfo {year} {2013})},\ \Eprint
  {http://arxiv.org/abs/1208.6415} {arXiv:1208.6415 [nucl-ex]} \BibitemShut
  {NoStop}%
\bibitem [{\citenamefont {Vary}\ \emph {et~al.}(2018)\citenamefont {Vary},
  \citenamefont {Basili}, \citenamefont {Du}, \citenamefont {Lockner},
  \citenamefont {Maris}, \citenamefont {Pal},\ and\ \citenamefont
  {Sarker}}]{Vary:2018jxg}%
  \BibitemOpen
  \bibfield  {author} {\bibinfo {author} {\bibfnamefont {J.~P.}\ \bibnamefont
  {Vary}}, \bibinfo {author} {\bibfnamefont {R.}~\bibnamefont {Basili}},
  \bibinfo {author} {\bibfnamefont {W.}~\bibnamefont {Du}}, \bibinfo {author}
  {\bibfnamefont {M.}~\bibnamefont {Lockner}}, \bibinfo {author} {\bibfnamefont
  {P.}~\bibnamefont {Maris}}, \bibinfo {author} {\bibfnamefont
  {S.}~\bibnamefont {Pal}}, \ and\ \bibinfo {author} {\bibfnamefont
  {S.}~\bibnamefont {Sarker}},\ }\href {\doibase 10.1103/PhysRevC.98.065502}
  {\bibfield  {journal} {\bibinfo  {journal} {Phys. Rev.}\ }\textbf {\bibinfo
  {volume} {C98}},\ \bibinfo {pages} {065502} (\bibinfo {year} {2018})},\
  \Eprint {http://arxiv.org/abs/1809.00276} {arXiv:1809.00276 [nucl-th]}
  \BibitemShut {NoStop}%
\bibitem [{\citenamefont {Wang}\ \emph {et~al.}(2019)\citenamefont {Wang},
  \citenamefont {Hayes}, \citenamefont {Carlson}, \citenamefont {Dong},
  \citenamefont {Mereghetti}, \citenamefont {Pastore},\ and\ \citenamefont
  {Wiringa}}]{WANG:2019}%
  \BibitemOpen
  \bibfield  {author} {\bibinfo {author} {\bibfnamefont {X.}~\bibnamefont
  {Wang}}, \bibinfo {author} {\bibfnamefont {A.}~\bibnamefont {Hayes}},
  \bibinfo {author} {\bibfnamefont {J.}~\bibnamefont {Carlson}}, \bibinfo
  {author} {\bibfnamefont {G.}~\bibnamefont {Dong}}, \bibinfo {author}
  {\bibfnamefont {E.}~\bibnamefont {Mereghetti}}, \bibinfo {author}
  {\bibfnamefont {S.}~\bibnamefont {Pastore}}, \ and\ \bibinfo {author}
  {\bibfnamefont {R.}~\bibnamefont {Wiringa}},\ }\href {\doibase
  https://doi.org/10.1016/j.physletb.2019.134974} {\bibfield  {journal}
  {\bibinfo  {journal} {Physics Letters B}\ }\textbf {\bibinfo {volume}
  {798}},\ \bibinfo {pages} {134974} (\bibinfo {year} {2019})}\BibitemShut
  {NoStop}%
\bibitem [{\citenamefont {Yao}\ \emph {et~al.}(2019)\citenamefont {Yao},
  \citenamefont {Bally}, \citenamefont {Engel}, \citenamefont {Wirth},
  \citenamefont {Rodríguez},\ and\ \citenamefont {Hergert}}]{Yao:2019rck}%
  \BibitemOpen
  \bibfield  {author} {\bibinfo {author} {\bibfnamefont {J.}~\bibnamefont
  {Yao}}, \bibinfo {author} {\bibfnamefont {B.}~\bibnamefont {Bally}}, \bibinfo
  {author} {\bibfnamefont {J.}~\bibnamefont {Engel}}, \bibinfo {author}
  {\bibfnamefont {R.}~\bibnamefont {Wirth}}, \bibinfo {author} {\bibfnamefont
  {T.}~\bibnamefont {Rodríguez}}, \ and\ \bibinfo {author} {\bibfnamefont
  {H.}~\bibnamefont {Hergert}},\ }\href@noop {} {\  (\bibinfo {year} {2019})},\
  \Eprint {http://arxiv.org/abs/1908.05424} {arXiv:1908.05424 [nucl-th]}
  \BibitemShut {NoStop}%
\end{thebibliography}%

\end{document}